\newcommand{\beq}{\begin{equation}}
\newcommand{\eeq}{\end{equation}}
\newcommand{\hg}{ }

\documentclass[aps,twocolumn,showpacs, showkeys,reprints]{revtex4-1}

\usepackage{amssymb}
\usepackage{hyperref}
\usepackage{epsfig}
\usepackage{natbib}
\usepackage{amsmath}
\usepackage{times}
\usepackage{psfrag}
\usepackage{braket}
\usepackage{textcomp}

\usepackage{algorithm}
\usepackage{algpseudocode}

\usepackage{graphicx}
\usepackage{subfigure}
\usepackage{float}
\begin{document}


\title{Stochastic first passage time accelerated with CUDA}

\author{Vincenzo Pierro}
\author{Luigi Troiano}
\author{Elena Mejuto}

\affiliation{Dept. of Engineering, University of Sannio, Corso Garibaldi, 107, I-82100 Benevento, Italy}

\author{Giovanni Filatrella }
\affiliation{Dept. of Sciences and Technologies, University of Sannio, Via Port'Arsa, 11, I-82100 Benevento, Italy}

\begin{abstract}

The numerical integration of stochastic trajectories to estimate the time to pass a threshold is an interesting physical quantity, for instance in Josephson junctions and atomic force microscopy, where the full trajectory is not accessible.
We propose an algorithm  suitable for  efficient implementation on graphical processing unit in CUDA environment. 
The proposed approach  for well balanced loads achieves almost perfect scaling with the number of available threads and processors, and allows an acceleration of about  
$400 \times$ with a GPU GTX980 respect to standard multicore CPU.
This method allows with off the shell GPU to challenge problems that are otherwise prohibitive, as thermal activation in slowly tilted potentials.
In particular, we demonstrate that it is possible to simulate the switching currents distributions of Josephson junctions in the timescale of actual experiments.

\end{abstract}



\maketitle

%%%%%%%%%%%%%%%%%%%%%%%%%%%%%%%%%%%%%%%%%%%%%%%%%%%%%%%%%%%%%%%%%%%%%%%%%%
\section{ Introduction}
\label{Introduction}
%%%%%%%%%%%%%%%%%%%%%%%%%%%%%%%%%%%%%%%%%%%%%%%%%%%%%%%%%%%%%%%%%%%%%%%%%%
%
% teoria e pratica importanza
%
The passage across a threshold of a stochastic process, the First Passage Time (FPT) process \cite{Risken89}, is a valuable tool to study theoretical properties of random systems
(e.g. Kramers'  rate theory \cite{Pollak16, Mazo10}). 
From the experimental point of view,  in Josephson Junctions (JJ) superconducting devices  \cite{Barone82} the measurement of the threshold for the escapes is the only possibility (or the simplest way) to gain information on the internal dynamics of systems as quantum devices \cite{DevoretBook,Pierro16}, threshold detectors \cite{Pekola04,Filatrella2010,Addesso2012} and arrays of JJ \cite{Gross13,Shukrinov17} of great interest for high frequency, up to the THz region, local oscillators \cite{Galin15}. 
The threshold for escapes is an essential physical information also in diverse systems{\hg . Examples are the atomic force spectroscopy (AFS) \cite{Freund09},  in some cases characterized by energy barrier similar to the washboard potential \cite{Mazo13} and the fast reversal of nanoparticles in a thermal bath \cite{Smirnov10}}. 

FPT process differs from standard stochastic evolution \cite{Januszewski10}, for the physically interesting quantity is no more some dynamical observable (e.g., the position of a particle) as a function of time, but the (random) time to reach a given coordinate.
Thus, while in the stochastic evolution one is interested in the probability distribution of the positions at a given time, in the analysis of passage times one deals with the probability distribution of the time necessary to reach a given position.
This is necessary when there exist special points where something happens that can be recorded by the instruments, as it is the case in the JJ physics that we discuss in this work.
This inversion of the roles of time and position calls for a different approach to parallel calculation, and prevents the straightforward application of already existing stochastic parallelization for the determination of FPTs.
Computationally efficient methods are advantageous in JJ physics for a number of reasons.
First, measurements are performed while changing the bias (that is, in the mechanical analogue, tilting the potential), and thus the resulting nonequilibrium problem requires numerical simulations.
Second, the bias change occurs on the scale of the conventional "DC" electronics, typically in the scale of kHz, while the characteristic time scale for the Josephson dynamics is in the range $0.1 \div 1$ THz, thus realistic and accurate simulations require something like $10^9 \div 10^{12}$ integrations steps.
Third, if JJ are to be used as signal detectors, the tests require extensive simulations to accurately retrieve the response to the binary hypothesis.
Fourth, JJ can be employed to form larger structures, or arrays {\hg and extended or long junctions where vortices are nucleated \cite{Fedorov09}}, that evidently increase the computational burden as many JJs are employed.
Fifth, as will be discussed in details in Sect. \ref{Sec:Arrhenius}, the estimate of the quasipotential \cite{graham85,kautz94} requires simulations for extremely rare escape events. 
However, all the above problems amount to the collection of random exit times; If stochastic replicas can be efficiently assigned to different processor units one straightforwardly benefits of an accelerated integration.
A possible method to achieve efficient distributed computation is the topic of this work.

The call for a specific algorithm for the numerical evaluation of FPTs can be summarized  in a rough physical and intuitive way as follows.
If the representative trajectory of a system is computed by a single processor (with sequential algorithms), in both stochastic evolution and random FPTs the processor just integrates the stochastic equation, and when the goal is reached (either the final position or the passage time, that are the desired information)
 the computation is arrested and a new calculation can begin.
Thus, in single processor calculations one can use with little changes the algorithm performing the integration of stochastic equations also for the computation of the FPTs.
Instead, for parallel computing the termination times on the processors are different, inasmuch the stochastic nature of the process entails different FPTs.
It is the lack of synchrony between the processors that limits the efficiency of parallel computation.
Our purpose is to device an efficient parallel algorithms that assigns a new job to the processor as soon as it has terminated its task, and to show how this can be efficiently done on a GPU in a CUDA environment.
{\hg There is a huge  literature for parallel solution of stochastic differential equations where an arrival time is to be calculated, see for instance \cite{Artemiev11} and references therein. 
The main point is that the performances are strongly related to the communication time, and hence to the processor architecture. 
Therefore, the knowledge accumulated for traditional parallel supercomputing cannot be directly transferred to CUDA environment. } 

The work is organized as follows. In Sect. \ref{physical} we discuss some physical examples where an accelerated (and cheap) procedure might prove useful, with special reference in Sect. \ref{Model} to the model equations for superconducting electronics.
The details of the computational parallel solution (CUDA oriented) are given in Sect. \ref{computational}, and the results of test simulations, together with the observed performances, are collected in Sect. \ref{Results}.
The conclusions are in the last Sect. \ref{Conclusions}.

%%%%%%%%%%%%%%%%%%%%%%%%%%%%%%%%%%%%%%%%%%%%%%%%%%%%%%%%%%%%%%%%%%%%%%%%%%
\section{The physical problem}
\label{physical} 
%%%%%%%%%%%%%%%%%%%%%%%%%%%%%%%%%%%%%%%%%%%%%%%%%%%%%%%%%%%%%%%%%%%%%%%%%%
To focus on a specific physical setting, we discuss a practical realization of the washboard potential, a superconductive JJ \cite{Barone82}. 
For this device, the FPT is of interest inasmuch it is associated to a measurable quantity. 
In JJ the trajectory it is not available because its measure is incompatible with the Heisenberg principle, for the intrinsic quantum nature of the representing coordinate.
What is experimentally accessible is the voltage, that is proportional to the velocity, associated to finite and large jumps, for the time scale in the THz region makes it prohibitive to follow the voltage time evolution.
The voltage jumps only occur when the phase passes the maximum of the metastable (washboard like) potential and enters the running regime, see Fig. \ref{fig:washboard}.
This exit across the separatrix is naturally formulated as a FPT on the top of the potential barrier and can be simulated with a GPU in CUDA environment.
Besides the general interest in FPT, there is an additional reason to pay a special attention to JJ. 
The  well established \cite{Buttiker83} noisy dynamics of Josephson junctions is of great contemporary interest 
\cite{Pekola04, Augello09, Groenbeck04, Groenbeck05} in connection with macroscopic quantum tunneling 
\cite{Martinis87,Shnirman97,Martinis02,Wallraff03b,Price10,Coskun12,Massarotti15,Makhlin01}. 
The quantum behavior of JJ makes them a good candidate for the realization of a quantum bit; however, the quantum regime is only observed after the ordinary thermal activated regime has been tamed \cite{Blackburn16}.
Thus, it is of interest to have precise numerical simulations of the thermal regime, simulations that are deeply connected with the theoretical predictions
(and corrections) of Kramers' formula  in the underdamped regime \cite{Mazo10}.

Finally, we wish to stress that there are other physical contexts where a similar setting is of practical interest.
For instance, AFS  is also due to the passage of the dip over the maximum of the atomic interaction potential \cite{Mazo13, Freund09}.
Interestingly, the model potential for AFS is often of the same type as the JJ, namely the washboard potential.
The description of the stochastic motion in a washboard potential is the subject of the next Section.

%%%%%%%%%%%%%%%%%%%%%%%%%%%%%%%%%%%%%%%%%%%%%%%%%%%%%%%%%%%%%%%%%%%%%%%%%%
\subsection{\label{Model} The washboard model: basic equations}

\begin{figure}[tbp]
\begin{center}
\includegraphics [scale=0.32]{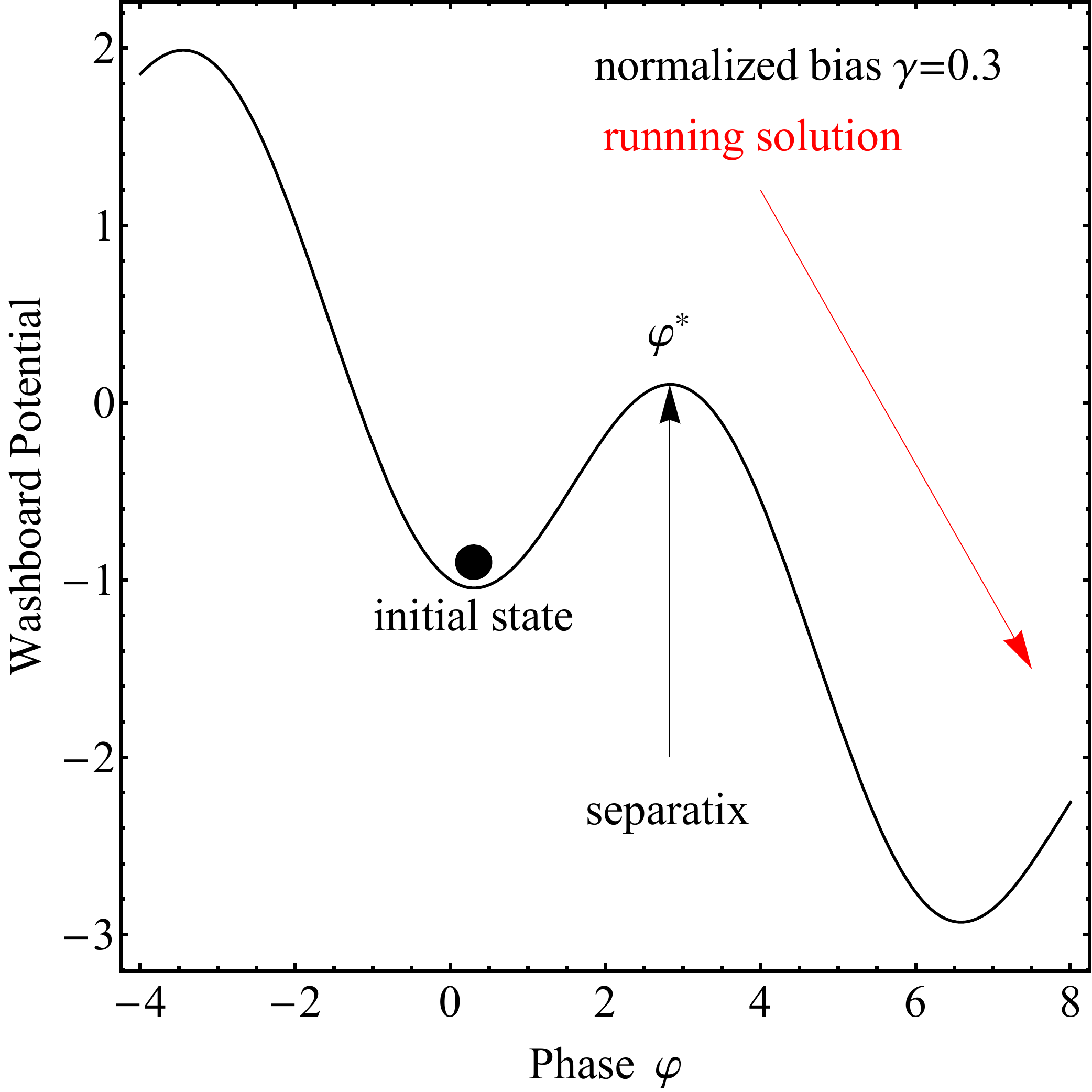}
\end{center}
\caption{(color online) Sketch of the washboard potential. 
The circle indicates the initial position of the virtual mechanical particle associated, for instance, to the quantum phase $\varphi$ of a JJ. 
The $\varphi^*$ value is the threshold indicating the separatrix.
The red arrow shows the running state of  the system after the escape over the barrier. 
Retrapping in subsequent minima never occurs in practice if the system is underdamped.}
\label{fig:washboard}
\end{figure}

%%%%%%%%%%%%%%%%%%%%%%%%%%%%%%%%%%%%%%%%%%%%%%%%%%%%%%%%%%%%%%%%%%%%%%%%%%
In this Section we describe the basic equations for the thermal stochastic washboard potential of Fig. \ref{fig:washboard}. 

%%%%%%%%%%%%___FIG___evolutions_____%%%%%%%
%%%%%%%%%%END FIGURE%%%%%%%%%%%%%%%%%%%%%%%%%%

To be specific, we derive the equations with reference to JJ. 
%%%%%%%%%%%%%%%%%%%%%%%%%%%%%%%%%%%%%%
%\subsection{Basic equations}
%\label{equations}
The usual model for a point-like JJ reads \cite{Barone82}:
\beq
\frac{C \hbar}{2e} \frac{d^2 \varphi}{d {t'} ^2} + 
\frac{1}{R}\frac{ \hbar}{2e} \frac{d \varphi}{d t'} + 
I_0 \sin \varphi  =
I_b + I (t')
\label{JJ}
\eeq
that includes inertia (determined by the capacitance $C$), dissipation (as governed by the dissipative element $R$) and fluctuations 
(the random current $I(t')$ supplied by the resistance), the nonlinear periodic term (the oscillating current of amplitude $I_0$), 
and a constant bias $I_b$.
Here, as usual, $\hbar$ is the reduced Planck constant, and $e$ is the electron charge. 
Fluctuations are assumed to be Gaussian with:
\begin{eqnarray}
\langle I(t') \rangle &=& 0,
\label{average} \\
\langle I(t') I(s') \rangle &=& \frac{2k_B T}{R}\delta(t'-s'),
\label{correlation}
\end{eqnarray}
where $k_B$ is the Boltzmann constant, $T$ is the absolute temperature, $\delta(\cdot)$ the Dirac function, $\langle \cdot \rangle$ is the expectation operator.
The normalized units \cite{Pierro16} are the following.
The current is normalized to the critical current $I_0$:
\beq
\label{curnorm}
\gamma = \frac{ I_b}{I_0}.
\eeq
The time is normalized to the quantity $\hbar/M$, where $M = C (\hbar/2e)^2 $ is the effective mass of the junction.
We introduce the time $\omega_N^{-1}= \hbar/E_C$, where $E_C$ is the Coulomb energy $E_C = (2e)^2/C$,  or the energy of the condenser charged by the elementary Cooper's pair. 
We have therefore:
\beq
t = t' \frac{E_C }{\hbar} =t' \frac{1}{C} \frac{(2e)^2}{\hbar}= t' \frac{\hbar}{M}. 
\eeq
Dissipation is given by the parallel resistor that in dimensionless units becomes
\beq
\label{beta}
\beta = \frac{1}{R} \frac{\hbar}{(2e)^2}.
\eeq
Also the fluctuating current is normalized to $I_0$: 
\beq
\gamma_N = \frac{ I(t')}{I_0}.
\eeq
The normalized noise  amplitude $D$ reads
\begin{eqnarray}
\label{WNAmp}
D  & = & \beta \frac{k_B T}{E_C}
\end{eqnarray}
that is, the thermal energy scaled to the energy $E_C$.
To evidentiate the fluctuation-dissipation theorem, one can also introduce the normalized dissipation $\beta$ and the normalized temperature $\theta$
in the fluctuation correlator (\ref{correlation}), we have:
\begin{equation}
\label{WNtheta}
\theta  =  \frac{ k_B T}{E_C} \Rightarrow  \nonumber 
D    =  \beta \theta.
\end{equation}

Using the above normalizations Eq. (\ref{JJ}) becomes
\beq
\frac{d^2 \varphi}{d {t} ^2} + \beta \frac{d \varphi}{d t} =  V_0 \left(-\sin \varphi + \gamma + \gamma_N\right),
\label{JJn}
\eeq
where $V_0= E_J/E_C$ and $E_J=I_0 \hbar/(2 e)$ is the energy barrier of phase particle.
The statistical properties of the Gaussian (thermal) noise normalized current $\gamma_N(t)$ are fully described by 
\begin{eqnarray}
\langle \gamma_N(t) \rangle &=& 0, 
\label{WNAverage} \\
\langle \gamma_N(t) \gamma_N(t+\Delta t) \rangle &=& 2D \, \delta \left (\Delta t \right)
= 2\beta \theta \, \delta \left (\Delta t \right) .
\label{WNCorrelation}
\end{eqnarray}
Eqs. (\ref{JJn},\ref{WNAverage},\ref{WNCorrelation}) constitute the physical setting in which we describe the method for accelerated computation of the FPTs.

%%%%%%%%%%%%%%%%%%%%%%%%%%%%%%%%%%%%%%%%%%%%%%%%%%%%%%%%%%%%%%%%%%%%%%%%%%
\subsection{The connection between escape times and switching currents probability distribution}

\begin{figure}[tbp]
\begin{center}
\includegraphics [scale=0.25, angle=0]{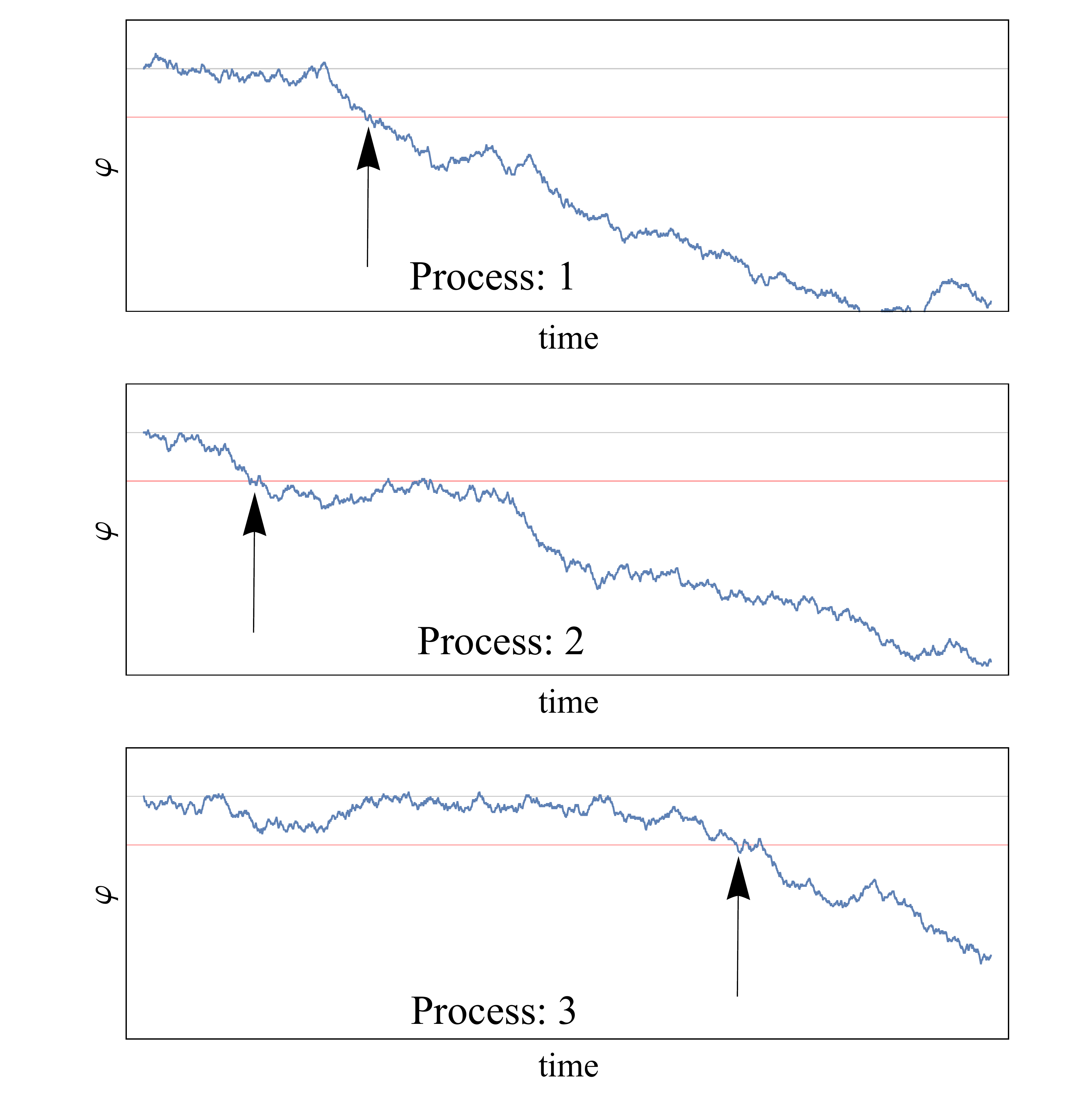}
\end{center}
\caption{(color online) Sketch of independent parallel stochastic evolutions, each belonging to a different process. 
The upper black horizontal line denotes the initial point, and the lower red line is the separatrix that defines the escapes. 
The claim that the FPT of a single realization has been computed occurs when the random trajectory touches the red line (i.e., the time instants determined by the black vertical arrows).}
\label{fig:evolution}
\end{figure}

%%%%%%%%%%%%%%%%%%%%%%%%%%%%%%%%%%%%%%%%%%%%%%%%%%%%%%%%%%%%%%%%%%%%%%%%%%
The evolution of a JJ according to the Langevin Eqs. (\ref{JJn},\ref{WNAverage},\ref{WNCorrelation}) is the basis to evaluate the exit time, that is the practically available physical  quantity.
Schematically, the trajectories that determine the  random exit times are represented in Fig. \ref{fig:evolution}.
The physically interesting quantity is the time at which the phase $\varphi$ hits the separatrix, because in the Josephson case this is the time at which a sizable voltage appears.
This sudden passage from a static metastable solution to a running solution across the separatrix $\varphi^{*}$, occurs at random times under the influence of noise. 
Rather than the direct measurement of the escape time, it is convenient in experimental setup to ramp the bias current, and to read the current at which the passage occurs.
Thus, the exit time (or the FPT) becomes a {\it Switching Current} (SC) {\hg if the reverse current probability is negligible \cite{Pankratov97}, which is commonly the case for underamped systems. 
Also, there is a deep connection, for overdamped systems, between the MFPT and the mean transition time, that is the statistical quantity that corresponds to the switching currents in JJ. 
However, in the first place, from a computational point of view the algorithms for the MFPT and the mean transition time are essential identical. In the second place, the MFPT are more general and of larger applications in systems outside the JJ physics. 
We also notice that, if mean transition times are to be used, one can use a modified Kramers formula \cite{Malakhov96}}.
The histogram of the so defined SCs represents the most important physical feature, from which a number of properties are deduced.
% Computationally-wise, to evaluate the FPT or the switching currents is equivalent, as will be shown in Sect. \ref{programming}
Computationally-wise, the evaluation of  the FPT or the switching currents is a rather similar task, as will be shown in Sect. \ref{programming}.

%%%%%%%%%%%%___FIG___evolutions_____%%%%%%%
%%%%%%%%%%END FIGURE%%%%%%%%%%%%%%%%%%%%%%%%%%
\section{GPU computation with CUDA}
\label{computational}
%%%%%%%%%%%%%%%%%%%%%%%%%%%%%%%%%%%%%%%%%%%%%%%%%%%%%%%%%%%%%%%%%%%%%%%%%%%%%%%%%%%%
We here demonstrate that it is possible to accelerate the computation of the FPTs, first describing the general structure in Sect. \ref{GPUarch} and the logic of the coding in Sect. \ref{programming}.
The last Sect. is devoted to the speed-up analysis.

\subsection{\label{GPUarch} GPU Architecture}

\begin{figure}
\center
\includegraphics [width=0.9\columnwidth]{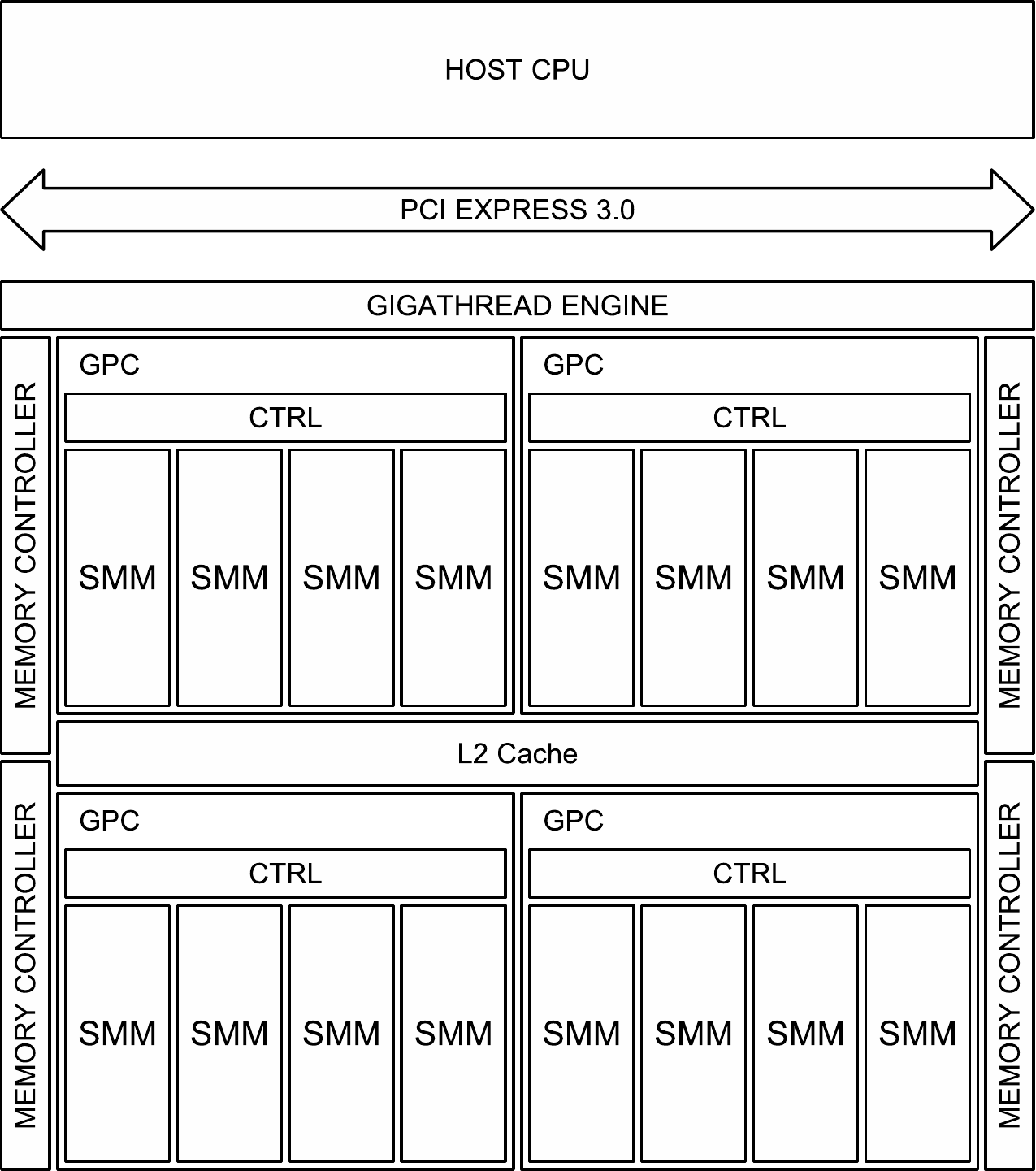}
\caption{Cores layout for the GPU routine. 
Here GPC stands for Graphic Processors Cluster and SMM for Maxwell Streaming Multiprocessors.}
\label{fig:GPU}
\end{figure}

As reference we use the NVIDIA Maxwell micro architecture used in the GEFORCE GTX980 graphics card based on GM204 GPU. This GPU, as shown in Fig.\ref{fig:GPU}, is composed of an array of 4 Graphics Processing Clusters (GPC), 16 Streaming Multiprocessors (SMM), and 4 memory controllers. 

In particular the GTX980 GPU is a 393 mm$^2$ die made of 5.2 billion transistors at 28nm integration scale, equipped with 2048 cores running at 1126 MHz (boost clock at 1216 MHz), able to reach 4.612 TFLOPs. 
The on-board memory is 4096MB, with L2 Cache size of 2048KB and memory clock at 7010 MHz, able to reach a bandwidth of 224.3 GB/sec.

The GPU architecture is based on an array of Graphics Processing Clusters (GPCs). These high-level blocks include hardware resources to perform almost all graphics processing, including pixel, texture, raster, vertex and geometry operations. 
In practice, the blocks are self-contained GPUs whose work is scheduled and coordinated by the 
GigaThread Engine at higher level. The blocks have access to memory by dedicated Memory Controllers. In each GPC, the processing is further parallelized using four Maxwell Streaming Multiprocessors (SMMs).

\begin{figure}
\center
\includegraphics [width=0.6\columnwidth]{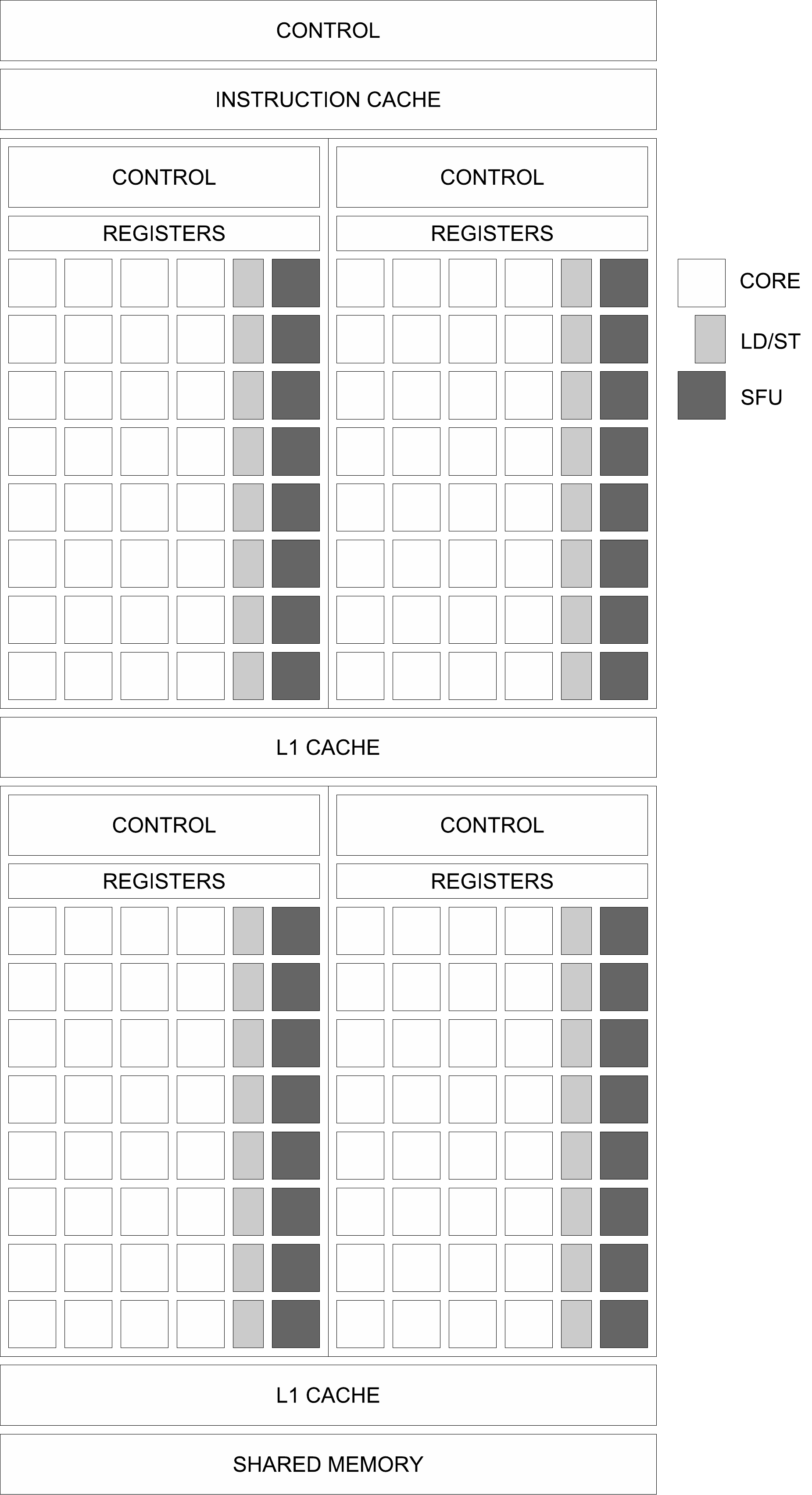}
\caption{Enlargment of the Maxwell Streaming Multiprocessors (SMM) blocks.}
\label{fig:SMM}
\end{figure}

\begin{figure}
\center
\includegraphics [width=0.4\columnwidth]{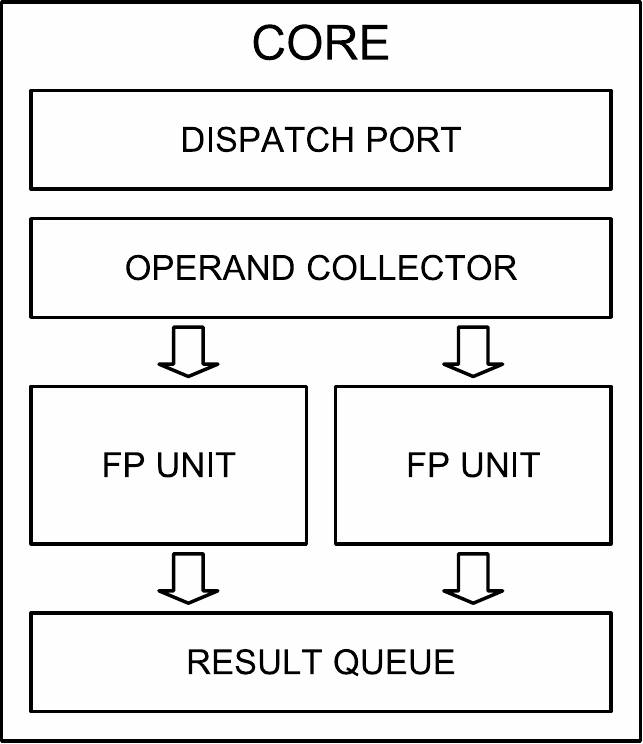}
\caption{Details of the CUDA core.}
\label{fig:core}
\end{figure}

As shown in Fig.\ref{fig:SMM}, each SMM features control logic, L2 cache memory and instruction cache memory, shared memory and four blocks. Each block has dedicated control logic and registers, that gain access to 32 CUDA cores organized in a grid.
Each row of four cores has access to load/store units to calculate source and destination addresses, to move (in/out) the data at each address and to cache or DRAM. 
In addition, for each row there is a Special Function Unit (SFU) designed to implement instructions such as sin, cosine, reciprocal, and square root.
The basic processing unit is a CUDA core (Fig.\ref{fig:core}). 
This computational unit is a simple scalar processor made of a fully pipelined integer arithmetic
logic unit (ALU) and floating point unit (FPU), based on IEEE 754-2008 floating-point standard, which improves the single and double precision multiplication and multiply-addition. The ALU is a 32-bit unit, optimized to support 64-bit operations. 
Among, the instructions there are Boolean, shift, move, compare, convert, bit-field extract, bit-reverse insert, and population count.

%%%%%%%%%%%%___FIG___3_____%%%%%%%

The algorithm we here propose for the CUDA architecture allows to reach a massive parallelization of simulations, as those depicted in Fig.\ref{fig:evolution}. 
Escape times are collected in a floating-point array $T_e$ (see the Algorithm \ref{algo:escape}) allocated to the GPU memory. 
The size of the array corresponds to the number of simulation runs one plans to execute. 
The code of each run is embedded in a \emph{kernel} and sent to the GPU using the host interface. The CUDA programming model is based on kernel, that is the portion of code that can be processed in parallel by threads. The kernel is fetched by the GigaThread Engine and sent to GPC and SMM.
Each run is assigned to a specific location in the array and scheduled to be executed as thread by a core. Threads are scheduled by each SMM in groups of 32 parallel threads called "warps". 
Each SMM features four warp schedulers and four dispatch units, so that four warps can be issued and executed concurrently. 
Each simulation run is executed independently from the other loads: when a core terminates a run, the next scheduled thread can be assigned to the core. 
It is therefore possible to exploit the hardware parallelism offered by the GPU almost to the full extent of its potentiality.

%%%%%%%%%%END FIGURE%%%%%%%%%%%%%%%%%%%%%%%%%%

%%%%%%%%%%%%___FIG___4_____%%%%%%%
%%%%%%%%%%END FIGURE%%%%%%%%%%%%%%%%%%%%%%%%%%

%%%%%%%%%%%%___FIG___5_____%%%%%%%
%%%%%%%%%%END FIGURE%%%%%%%%%%%%%%%%%%%%%%%%%%

\begin{algorithm}[H]
\caption{Kernel used for the simulations of the escape time)}
\label{algo:escape}
\begin{algorithmic}[1]
\Procedure{EscapeTime}{$V_0$, $\beta$, $\gamma$, $D$, $h$}

\Statex $V_0$, normalized potential barrier
\Statex $\beta$, friction coefficient
\Statex $\gamma$, normalized bias current
\Statex $D$, Gaussian noise amplitude
\Statex $h$, Euler time step

\Statex

\State $pos \gets cuda.grid(1)$
\If {$pos < sizeof(T_e)$}

\State $\sigma \gets \sqrt{2D \cdot h}$
\State $\varphi_0 \gets \arcsin{\gamma}$
\State $th \gets \pi - \varphi_0$
\State $\dot\varphi_0 \gets 0$, initial phase velocity

\State $k \gets 1$ 
\While{$k \leq klim$}
	\State $r \gets nrand()$
    \State $\varphi_{k} \gets \varphi_{k-1} + \dot\varphi_{k-1} \cdot h$
    \State $\dot\varphi_k \gets \dot\varphi_{k-1} + (-\beta \dot \varphi_{k-1} - V_0 \sin \varphi_{k-1} + V_0 \gamma) \cdot h + \sigma \cdot r$
    
    \If{ $\varphi_k > th$}
         \State $T_{e}[pos] \gets k\cdot h$
         \State \Return
	\EndIf
\State $k \gets k+1$
    
\EndWhile
\State $T_{e}[pos] \gets -1$
\EndIf

\Statex
\EndProcedure

\end{algorithmic}
\end{algorithm}

\subsection{\label{programming} Programming}

\begin{figure}
\center
\includegraphics [width=0.8\columnwidth]{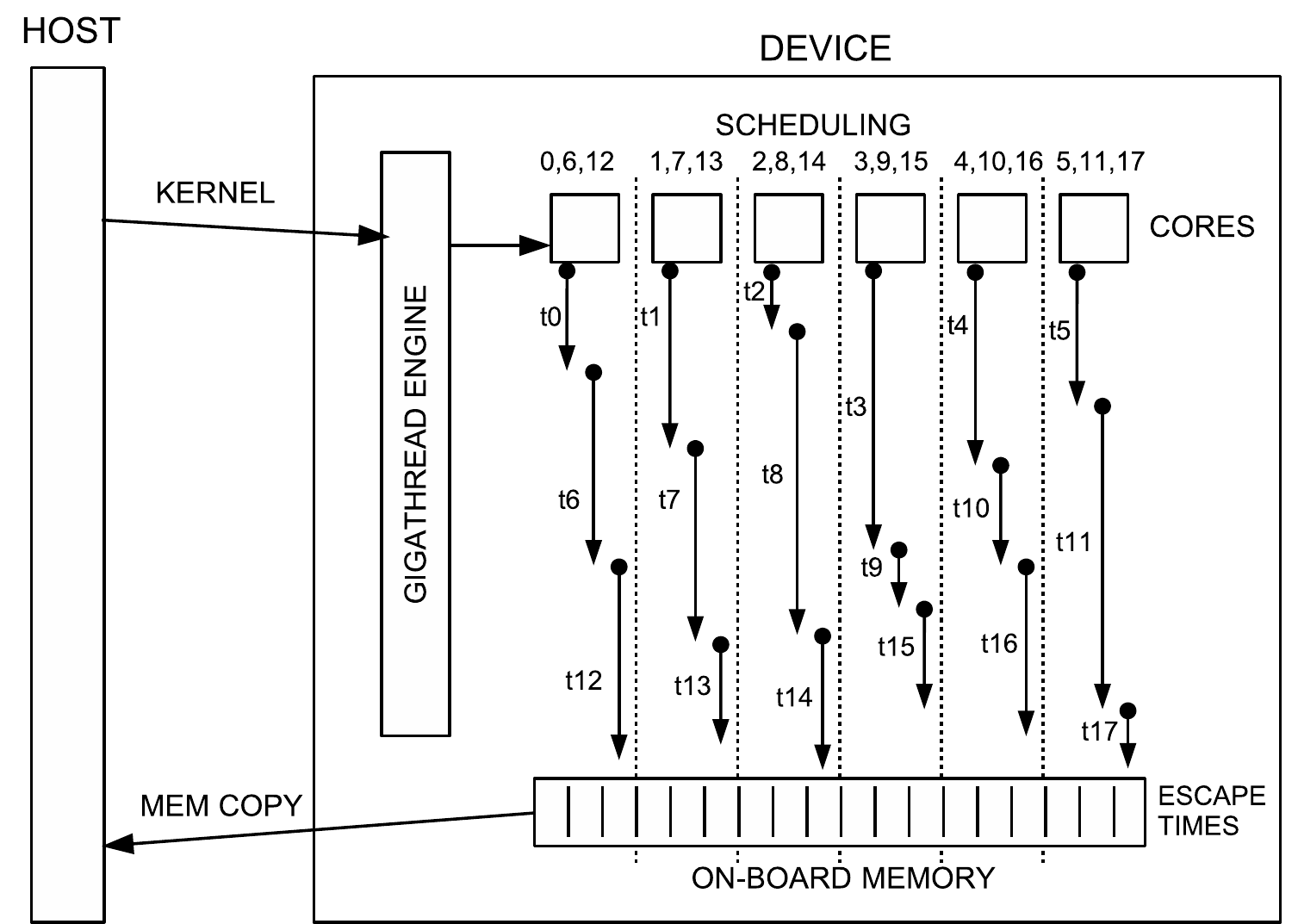}
\caption{Sketch of the streaming process, the GigaThread Engine that dictates the task scheduling. 
The processor computes an element ( e.g. $t0$, $t6$, ...) of the escape time sequence and eventually stores the result in the array $T_e$. 
The order of the single tasks termination does not correspond to the array index order of $T_e$. }
\label{fig:streaming}
\end{figure}

The simulation is coded as kernel in the environment CUDA. 
The pseudo-code of the kernel for the computation of the exit time distribution is outlined in Algorithm \ref{algo:escape}. 
The Algorithm is implemented by the \textproc{EscapeTime} procedure that accepts five arguments as input.
The symbols and the physical meaning of the first four parameters have been introduced in Eq. (\ref{JJn}).
The last parameter, the time step  $h$ of the stochastic integration method is used in an Euler scheme \cite{EulerInt}, lines $11$, $12$.
The simulations are retrieved following the scheme outlined in Figs. \ref{fig:evolution} and \ref{fig:streaming}, by integration of the stochastic differential equation (\ref{JJn}) and checking for the threshold crossing, see line $13$.
The sequence of operation in the escape time Algorithm \ref{algo:escape} is as follows
\begin{enumerate}
\item Each processor device begins the trajectory  integration (see Fig. \ref{fig:evolution} ) checking at each time step   if a threshold crossing (i.e., a FPT) has occurred
\item When the FPT has been encountered the integration ends on a processor device and the device communicates the escape time to the host processor  (see Fig. \ref{fig:streaming})
\item The host device assigns a new job to the processor device that has just completed the task
\end{enumerate}
In Algorithm \ref{algo:escape} the function $nrand$ at line $10$ generates a Gaussian random variable with zero average and unit variance.
This  Gaussian random generator is based on the Box-Muller transform \cite{Numrecip} applied to pseudo-random number obtained with the Mersenne Twister generator \cite{Matsumoto98}.
The corresponding CUDA core pseudocode is outlined in Algorithm \ref{algo:mtrg}. 
The computation is based on an internal subsequence that is initialized according to the \textproc{MTRG-Init} procedure. 
This subsequence is named $mt$ and it is made of 624 32-bits integers locally allocated to the device (line 2) and stored at L1 cache as shown in Fig.\ref{fig:SMM}. 
The seed, used as the first element of the sequence (line 3), is a unique value assigned to each thread.
This initial value is externally provided from the host code and passed to the kernel. 
To generate the seed we generally use the current time in milliseconds.

\begin{algorithm}[H]
\caption{Marsenne Twister Random Generator}
\label{algo:mtrg}
\begin{algorithmic}[1]

\Procedure{MTRG-Init}{seed}
\Statex $seed$, random sequence initial value
\Statex {}
\State $mt \gets cuda.local.array(624, dtype=int32)$
\State $mt[0] \gets seed$
\For { $i = 1 .. 623$}
	\State $m = (mt[i - 1] \wedge (mt[i - 1] >> 30)) + i)$
	\State $mt[i] = (\mathtt{1812433253} \cdot m) \wedge \mathtt{0xFFFFFFFF}$
\EndFor
\State \Return {$mt$}
\Statex
\EndProcedure
\Statex
\Procedure{MTRG-Extract}{$r$, $mt$}
\Statex $r$, sequence index 
\Statex $mt$, internal sub-sequence
\Statex {}
\State $ri \gets r \bmod 624$

\If { $ri = 0$}

	\For {$i = 0..623$}
        \State $y \gets mt[i] \wedge \mathtt{0x80000000}$
        \State $y \gets y + (mt[(i + 1) \bmod 624] \wedge \mathtt{0x7fffffff}))$
        \State $mt[i] \gets mt[(i + 397) \bmod 624] \oplus (y >> 1)$
        \If {$y$ is odd}
            \State $mt[i] \gets mt[i] \oplus \mathtt{0x9908b0df}$
		\EndIf
    \EndFor
\EndIf

\State $y \gets mt[ri]$
\State $y \gets y \oplus (y >> 11) \wedge \mathtt{0xFFFFFFFF}$
\State $y \gets y \oplus (y << 7) \wedge \mathtt{0x9D2C5680}$
\State $y \gets y \oplus (y << 15) \wedge \mathtt{0xEFC60000}$
\State $y \gets y \oplus (y >> 18) \wedge \mathtt{0xFFFFFFFF}$
    
\State \Return $y$

\Statex
\EndProcedure

\end{algorithmic}
\end{algorithm}

The seed used by \textproc{MTRG-Init} is obtained by adding the thread identification number to initial value. 
With this expedient each thread makes use of a new seed based on the assigned simulation run number. 
The scheme allows each thread to keep a unique independent pseudorandom integer 
sequence that is controlled by the initial seed. From the value $mt[0]$, all the following values of the subsequence $mt$ are obtained by iteration (lines 4-7). 
Once completed, the subsequence $mt$ is returned as a result of the initialization (line 8) and used by \textproc{MTRG-Extract} according to the sequence index $r$ and $mt$ (line 10).
The procedure works by blocks of 624 values.
Which value of the block to use is simply indexed by a modulo operation (line 11). 
At the beginning, and at each time they are employed (line 12 of Algorithm \ref{algo:escape}), a new block of values is computed from the previous one (lines 13-20). 
Then, the value $mt[ri]$ is transformed (lines 22-26 of Algorithm \ref{algo:mtrg}) and returned (line 27).  

\begin{algorithm}[H]
\caption{Kernel used for simulations of the switching currents}
\label{algo:switching}
\begin{algorithmic}[1]
\Procedure{SwitchingCurrent}{$\delta \gamma$, $V_0$, $\beta$, $D$, $h$}

\Statex $\delta \gamma$, normalized ramp step
\Statex $V_0$, normalized potential barrier
\Statex $\beta$, friction coefficient
\Statex $D$, Gaussian noise amplitude
\Statex $h$, Euler time step

\Statex

\State $\gamma_0 \gets 0$, initial normalized bias current

\State $pos \gets cuda.grid(1)$
\If {$pos < size of(\Gamma_e)$}

\State $\sigma \gets \sqrt{2D \cdot h}$
\State $\varphi_0 \gets \arcsin{\gamma}$
\State $th \gets \pi - \phi_0$
\State $\dot\varphi_0 \gets 0$, initial phase velocity

\State $k \gets 1$ 
\State $\gamma_k \gets \gamma_0$ 

\While{$\gamma_k \leq 1$}
	\State $r \gets nrand()$
	\State $\gamma_k \gets \gamma_0 + \delta \gamma \cdot k$
    \State $\varphi_{k} \gets \varphi_{k-1} + \dot\varphi_{k-1} \cdot h$
    \State $\dot\varphi_k \gets \dot\varphi_{k-1} + (-\beta \dot \varphi_{k-1} - V_0 \sin \varphi_{k-1} + V_0 \gamma_k) \cdot h + \sigma \cdot r$
    
    \If{ $\varphi_k > th$}
         \State $\Gamma_{e}[pos] \gets \gamma_k$
         \State \Return
	\EndIf
\State $k \gets k+1$
    
\EndWhile
\State $\Gamma_{e}[pos] \gets 1$
\EndIf

\Statex
\EndProcedure

\end{algorithmic}
\end{algorithm}

In conclusion the code \textproc{MTRG-Extract} returns a random integer uniformly distributed in the range of 32-bit integers.
This integer with a suitable linear normalization generate an uniform deviate in the range $[0,1]$. 
The uniform deviate is used in the external function $nrand$ with the Box-Muller transform to generate Gaussian pseudo random numbers with zero average and unit variance.

The proposed Algorithm \ref{algo:escape} is useful for the computation of escape time distribution by means of direct event simulations. 
In threshold device based on JJ it is experimentally simpler to retrieve the SCs distribution \cite{Fulton74}. 
The proposed code is straightforwardly changed to compute the SCs instead of the exit time, as per the \textproc{SwitchingCurrent} that is only slightly different from the escape time process (see Algorithm  \ref{algo:switching}).
During the computation, the bias current is increased from zero to $1$ (that in our normalized units, see Eq.(\ref{curnorm}),  corresponds to the maximum superconducting current that can flow through the JJ). 
The numerical method to increase the bias current at each time step, however, requires a particular care.
In fact the GPUs normally work in single precision.
When this is the case, the smallest bias increment is $10^{-7}$, that is the reciprocal of the largest integer in single precision. 
Consequently, for the extremely long simulations reported in the experiments demand that the bias current is not incremented with a single precision integer loop. 
The difficulty can be overcome with a tuple of $n$ 32-bit integers.

%%%%%%%%%%%%___FIG___6_____%%%%%%%
%%%%%%%%%%END FIGURE%%%%%%%%%%%%%%%%%%%%%%%%%%

%%%%%%%%%%%%%%%%%%%%%%%%%%%%%%%%%%%%%%%%%%%%%%%%%%%%%%%%%%%%%%%%%%%%%
\subsection{\label{parallelism} Parallelism and Speedup tests}
%%%%%%%%%%%%%%%%%%%%%%%%%%%%%%%%%%%%%%%%%%%%%%%%%%%%%%%%%%%%%%%%%%%%%

\begin{figure}[ht]
\center
\includegraphics [scale=0.25]{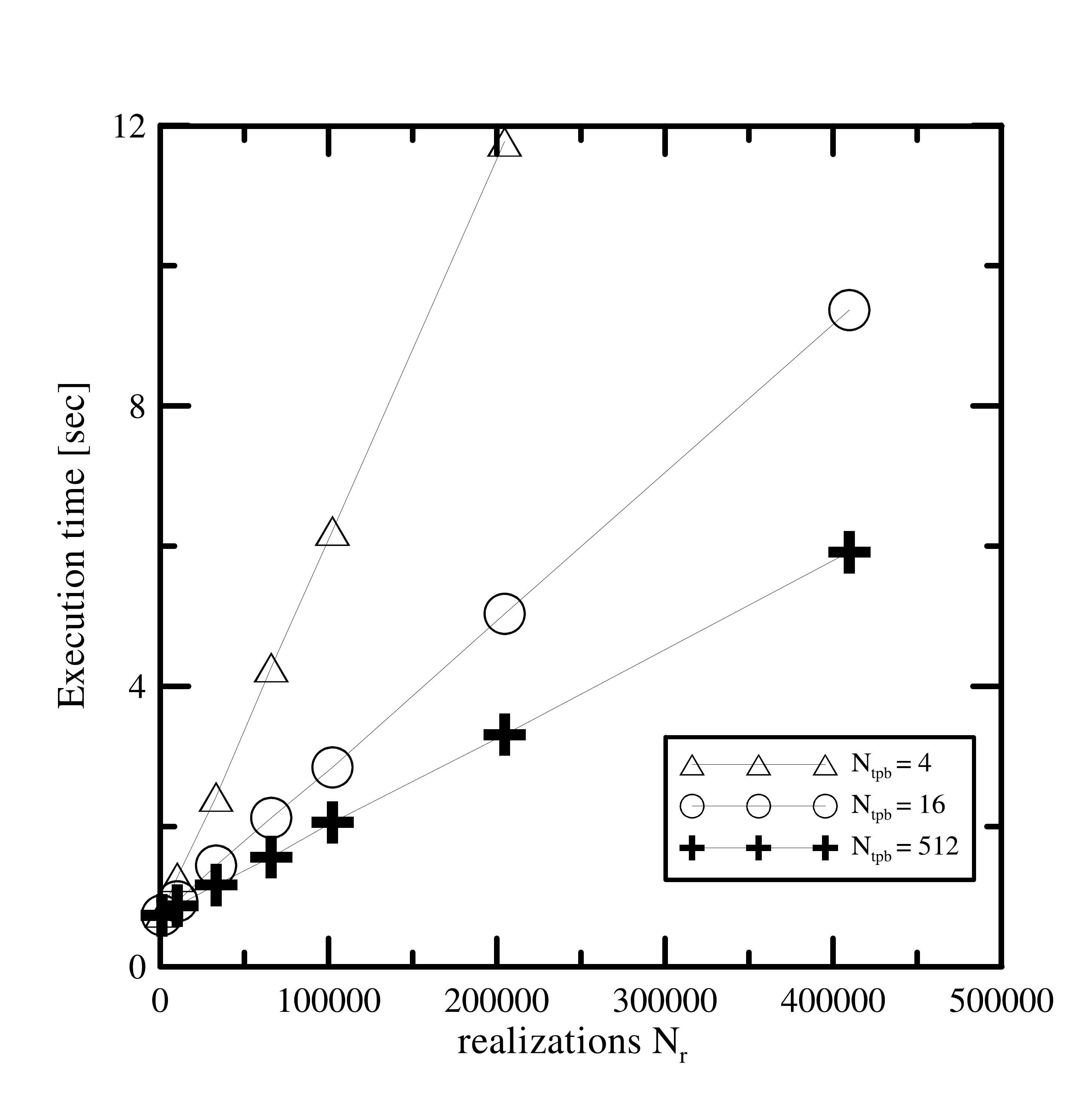}
\caption{Sketch of the execution time (in seconds) as a function of the number of realizations $N_r$ 
with different numbers of active threads per block $N_{tpb}$.}
\label{fig:exec}
\end{figure}

To investigate the scaling with the number of processing units, we have studied the execution time as a function of the number of 
realizations $N_r$ for three different numbers (denoted as $N_{tpb}$) of {\em threads per block} (TPB).
This is shown in Fig. \ref{fig:exec}.
From the Figure it is evident that for a given number of processors the execution time linearly scales with the number of realizations, as expected for serial computation.
The advantage is in the slope, that decreases when the number of units is increased.

%%%%%%%%%%%%___FIG___realizations_____%%%%%%%
%%%%%%%%%%END FIGURE%%%%%%%%%%%%%%%%%%%%%%%%%%

\begin{figure}[h]
\includegraphics [scale=0.25]{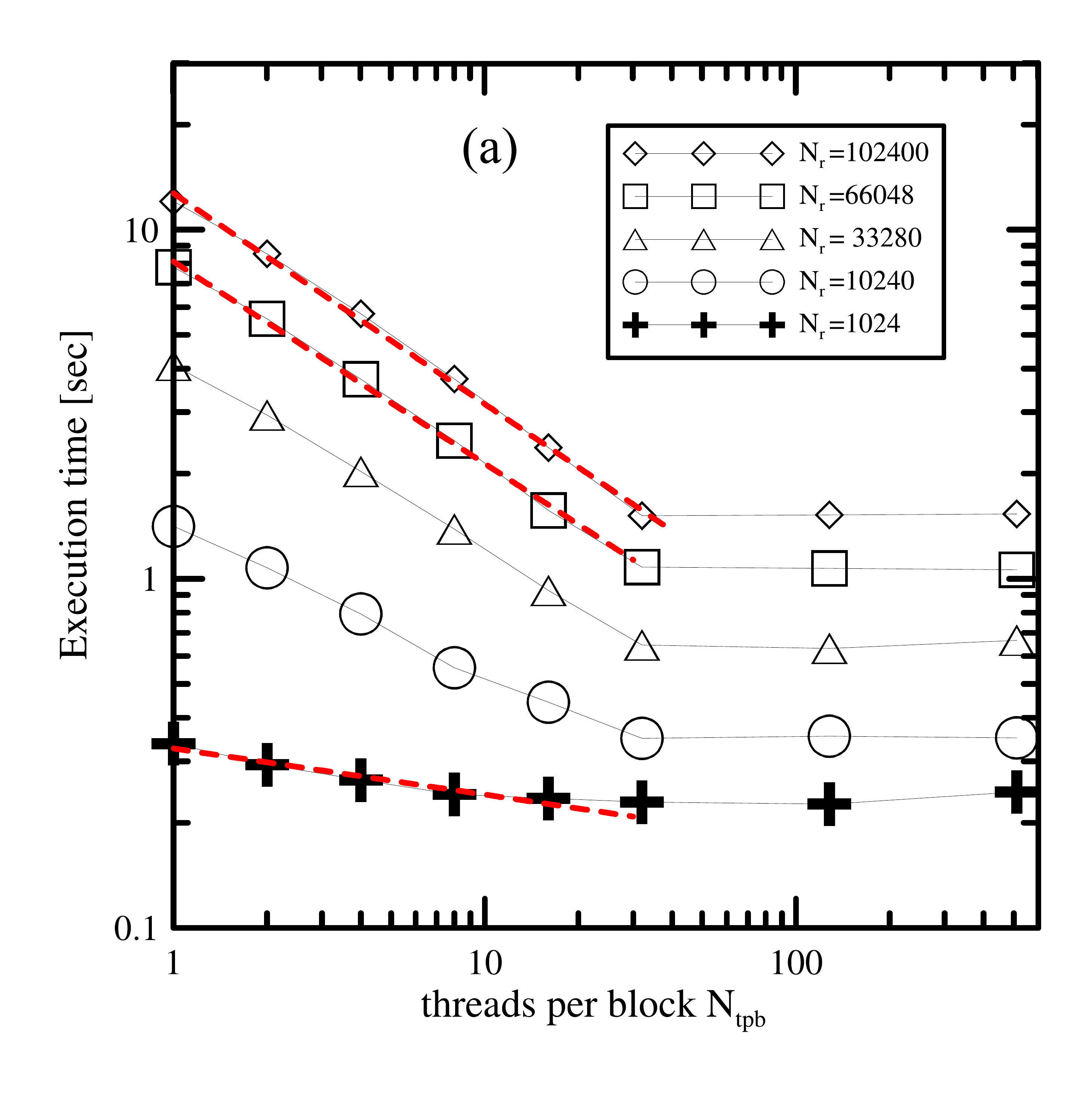}
\includegraphics [scale=0.25]{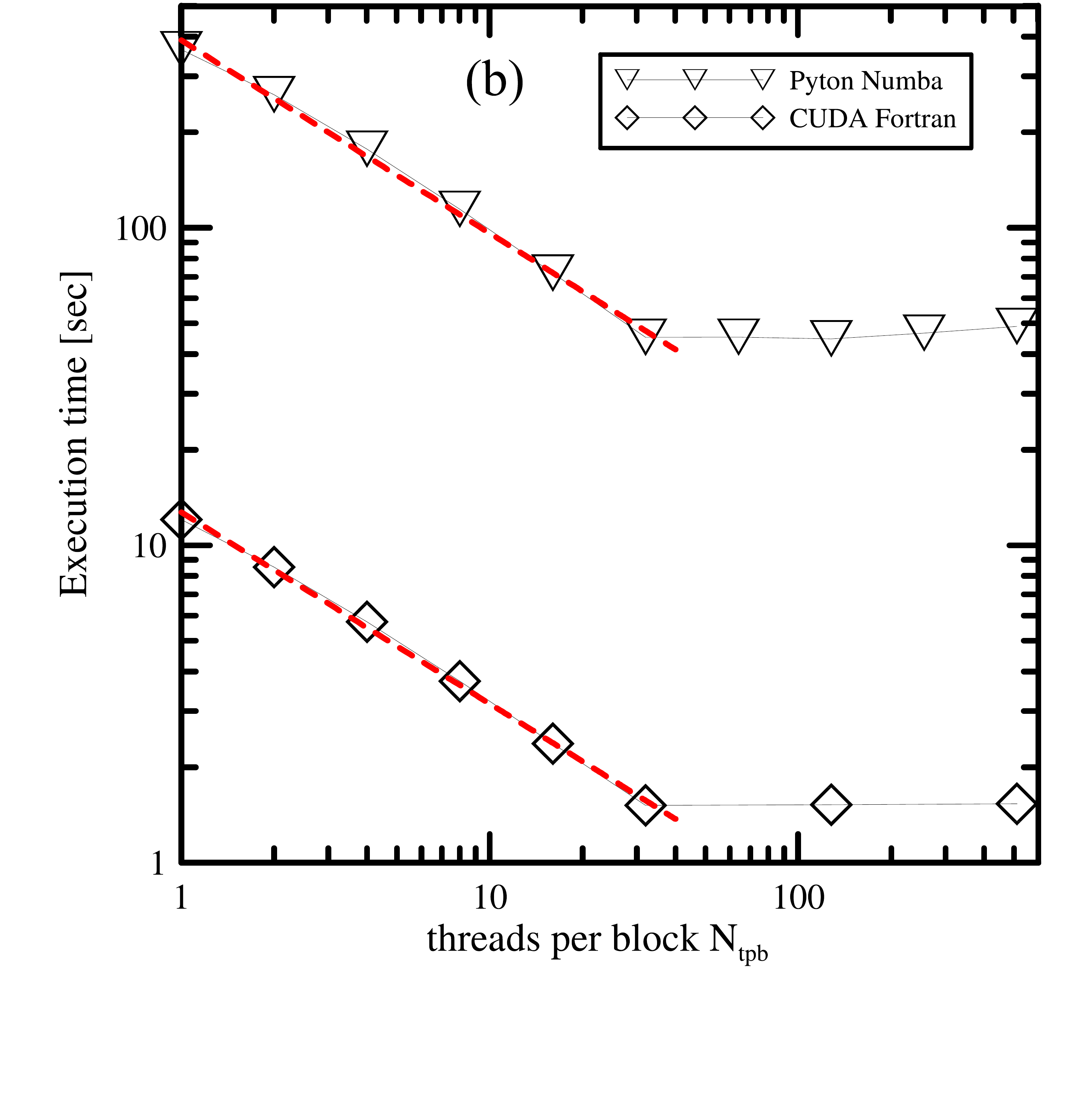}
\caption{
(color online) (a) Sketch of the execution time (in seconds) as a function of the number of TPB for different number of realizations $N_r$ reported in the legend.
The dashed red curves give the power law interpolation. For $N_r=1024$ the scaling coefficient is $0.13$, it converges for larger values of $N_r$ to the values $\approx 0.6$.
(b) Python Numba vs CUDA Fortran execution time (in seconds) as function of TPB. We note the same power law scaling (i.e. $\approx 0.6$), as a function of $N_{tpb}$ in both  practical implementations. 
}
\label{fig:exec_time_comparison}
\end{figure}

The advantage of a large number of processing units is shown in more details in Figs. \ref{fig:exec_time_comparison}(a,b).
In (a) we display the execution time as a function of the TPB for different numbers of realizations $N_r$, from $\simeq 10^3$ to $\simeq 10^5$. 
The figure essentially confirms the linear behavior of Fig. \ref{fig:exec}: the increased number of processor units proportionally scales with the required CPU time only above a certain number of threads. 
This number, that depends upon the number of realizations, represents the point at which the additional processors do not contribute.
In Fig. \ref{fig:exec_time_comparison}(b) we compare two technologies (Python Numba vs CUDA Fortran solution) to implement the same algorithm.
It is evident that the overall behavior is the same for both technologies, albeit the CUDA Fortran is much more efficient,  of about a factor $30$.
This is to be expected, as Numba is based on CUDA Toolkit version 7.5 which does not provide a native implementation of the random generator library, made available in version 8 as \texttt{CURAND\_DEVICE} library and used by the CUDA Fortran compiler. 
Thus, the execution of \textproc{MTRG-Extract} in Algorithm \ref{algo:mtrg} significantly impacts on the simulation kernel if the number of operations required by the random generator (MTRG and Box-Muller) is comparable to the number of operations for the integration. 
In addition, CUDA Fortran uses compiler optimizations that are not available to the LLVM compiler infrastructure on which Numba is based. In both figures we observe that time required by simulations decreases linearly with respect to number of active TPB. A larger number of threads improves the GPU occupancy, up to the limit of 512 TPB. 

%%%%%%%%%%%%___FIG___8_____%%%%%%%
%%%%%%%%%%END FIGURE%%%%%%%%%%%%%%%%%%%%%%%%%%

\begin{figure}
\center
\includegraphics [scale=0.28]{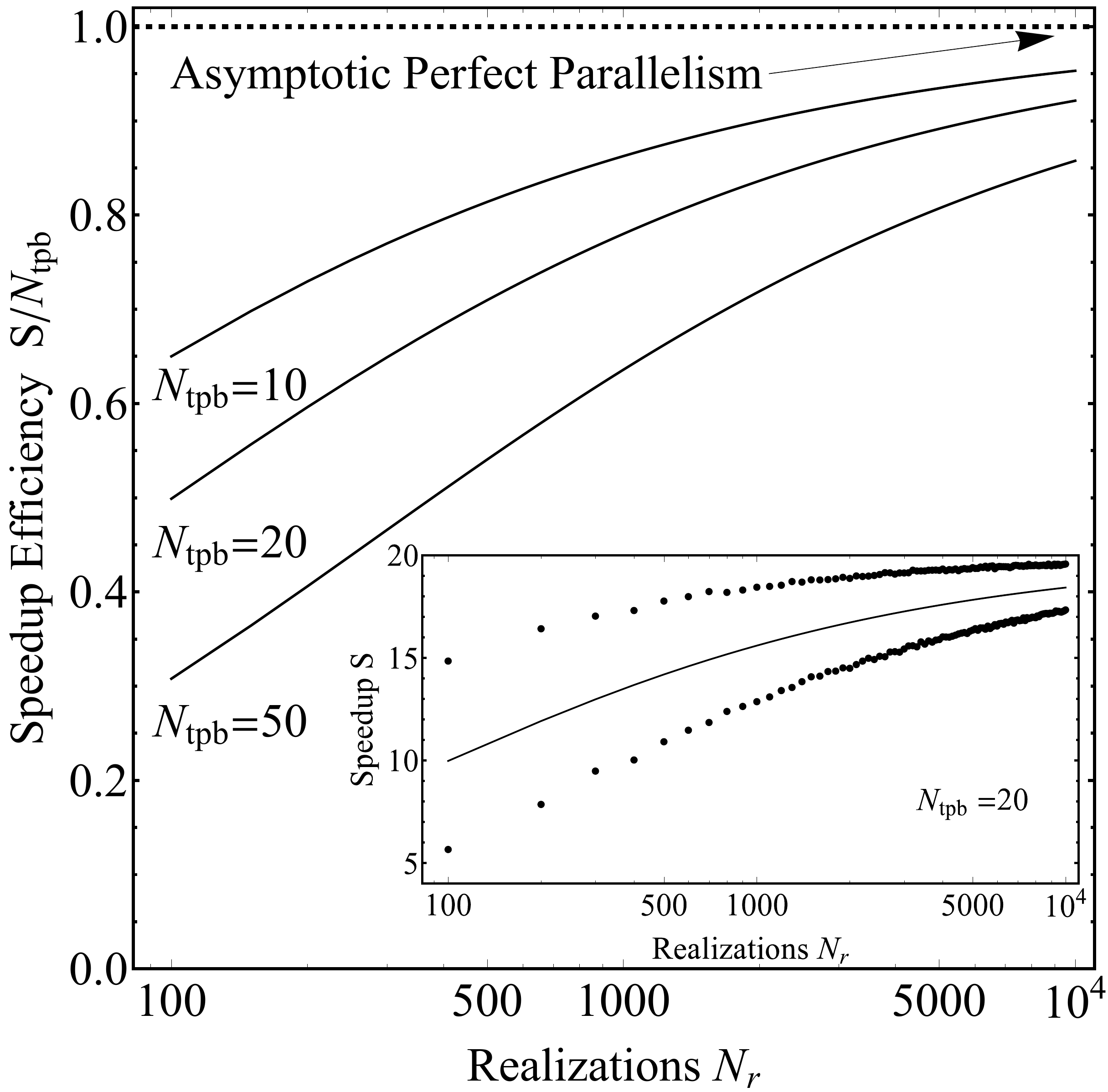}
\vspace{-0cm}
\caption{Speedup efficiency $S/N_{tpb}$ as a function of the number of realizations (on a log scale), as per Appendix A. 
The inset shows the speedup $S$ (solid line) for the particular case $N_{tpb}=20$, together with the $\pm 3\sigma$ uncertainty (dots) due to statistical variability.} 
\label{fig:speedup}
\end{figure}

The speedup efficiency, defined as the ratio $S/N_{tpb}$, is displayed in Fig. \ref{fig:speedup}.
In the efficiency the speedup $S$ is defined as the ratio between the average execution time for the sequential and the parallel algorithm
(see Eq.(\ref{eq:speedup}) and the Appendix A for the details of the model and notation). 
The data show that the asymptotic efficiency converges to $1$ as the number of realizations increases. 
For a constant number of realizations $N_r$, the efficiency makes worse increasing the number of threads per block (i.e., reducing the computation burden per thread $N_L$). 
This effect is manly due to the fluctuations in the execution time of each thread, that emerge in the small sample limit when the load is distributed over too many threads.

In the inset of Fig. \ref{fig:speedup} we show a  $3\sigma$ error limits due to the statistical variability of the speedup $S$ for the particular case $N_{tpb}=20$.
%%%%%%%%%%%%___FIG___   Appendix  _____%%%%%%%
%%%%%%%%%%END FIGURE%%%%%%%%%%%%%%%%%%%%%%%%%%
We remind that the statistical model employed in the Appendix neglects the short time behavior of the system.
However, the qualitative asymptotic behavior of the speedup, and more specifically of the statistics concerning the execution time, is mainly due to the exponential tail of the random variable distribution.

%%%%%%%%%%%%%%%%%%%%%%%%%%%%%%%%%%%%%%%%%%%%%%%%%%%%%%%%%%%%%%%%%%%%%%%%%%%%%%%%%%%%
\section{\label{Results} Tests on physical systems}

%%%%%%%%%%%%%%%%%%%%%%%%%%%%%%%%%%%%%%%%%%%%%%%%%%%%%%%%%%%%%%%%%%%%%%%%%%%%%%%%%%%%
At  the end of the previous Section we have investigated the execution time statistics of the Algorithms \ref{algo:escape} and \ref{algo:switching}.
In the following we collect numerical results of the benchmark physical system, that is of superconducting  JJs.
We do so to validate the whole method, for the CUDA arithmetic is poorer than the ordinary CPU arithmetic, in that it is single precision and with a lower level of fidelity \cite{floating}.

\subsection{\label{Sec:Arrhenius} Arrhenius plots}

\begin{figure}
\includegraphics [scale=0.35]{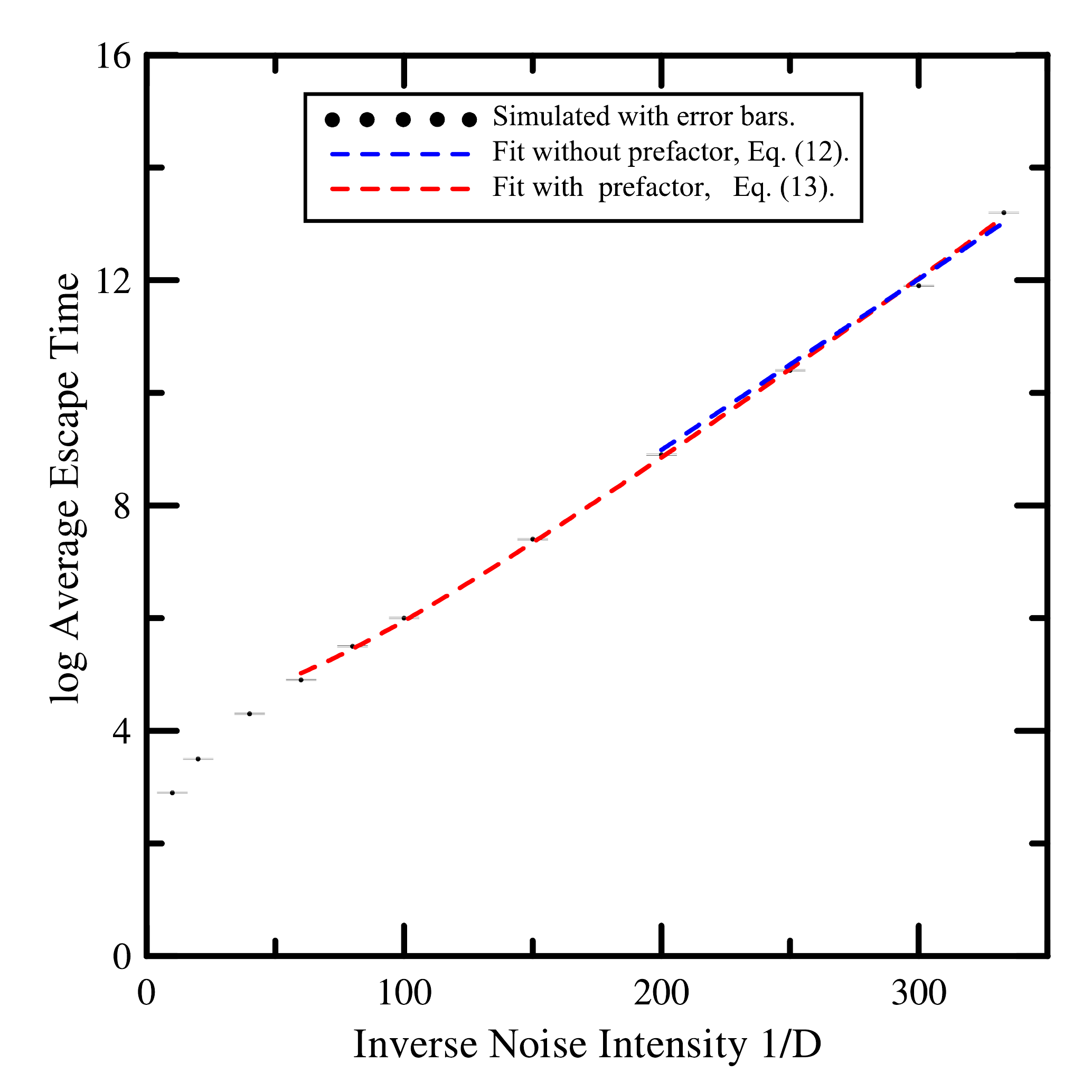}
\caption{(color online) Log of the average FPT (black dots) as a function of the inverse normalized temperature $D=\beta\theta$, compared to the theory, Eq. (\ref{Arrhenius}) 
(short dashed blue) and Eq. (\ref{kramers}) (long dashed red).
Parameters of the simulations are: $\gamma=0.5$, $\alpha =0.05$, $N_{r}=5120$, integration step $h=0.004$
The theoretical value for the energy barrier reads $\Delta U = 0.0342$, while the numerical evaluation reads $\Delta \tilde{U} \simeq 0.0303$ for fitting with Eq.(\ref{Arrhenius}) and $\Delta \tilde{U} \simeq 0.0359$ for fitting with Eq.(\ref{kramers}). The accuracy of the numerical method is $11\%$ and $4\%$, respectively. 
}
\label{fig:kramers_time}
\end{figure}

First, to validate the algorithms we have investigated the case of a constant bias JJ subject to different noise intensity, see Fig. \ref{fig:kramers_time}. 
In the Figure the error bar is not visible, for the speed of the algorithm has allowed to collect a large number of realizations ($N_r=5120$).
This high accuracy allows to distinguish between the Arrhenius behavior
\beq 
\label{Arrhenius}
\langle \tau \rangle \propto \exp{\frac{\Delta U}{\Delta \theta}}
\eeq
and the detailed Kramers rate, obtained taking into account the prefactor corrections, for the special case of moderately underdamped systems.
The model fits in this case with prefactor becomes \cite{Risken89}:
\beq 
\label{kramers}
\langle \tau \rangle \propto \frac{1}{\theta}\exp{\frac{\Delta U}{\Delta \theta}}.
\eeq
The accurate simulations of the mean FPT as a function of the inverse of the temperature is a task useful in several applications, most importantly to retrieve 
the so-called quasipotential (or pseudopotential) for non-Hamiltonian systems \cite{graham85,kautz94}.
When a bona fide potential does not exist or is not explicitly known, one can estimate an effective quasipotential from numerical simulations as those of Fig. \ref{fig:kramers_time}, reversing the logic of Eq. (\ref{kramers}), i.e., assuming:
\beq
\label{quasipotential}
\Delta U \equiv \frac{\log{(\theta \langle \tau \rangle)}}{\Delta \theta}.
\eeq
The resulting  method requires to simulate the system in the low noise limits, when the relations (\ref{Arrhenius},\ref{kramers}) are strictly valid \cite{graham85,kautz94}. 
By the same token, in this limit simulations are extremely long, and hence the call for fast, parallel simulations.
In this work we retrieve the potential $\Delta U$ form the slope of the Arrhenius plots of Fig. \ref{fig:kramers_time} of a system such as the Josephson potential associated to Eq.(\ref{JJn}).
In this case, being the potential known, we can use the analytic result to ensure that numerical simulations are reliable.
From the data of Fig. \ref{fig:kramers_time} we conclude that Kramers fitting offers a good estimate of the potential barrier $\Delta U$, with a relative error of few percents.

%%%%%%%%%%%%___FIG___   kramers_time  _____%%%%%%%
%%%%%%%%%%END FIGURE%%%%%%%%%%%%%%%%%%%%%%%%%%

%%%%%%%%%%%%%%%%%%%%%%%%%%%%%%%%%%%%%%%%%%%%%%%%%%%%%%%%%%%%%%%%%%%
\subsection{Escapes and switching current distributions}

\begin{figure}[H]
\includegraphics [scale=0.27]{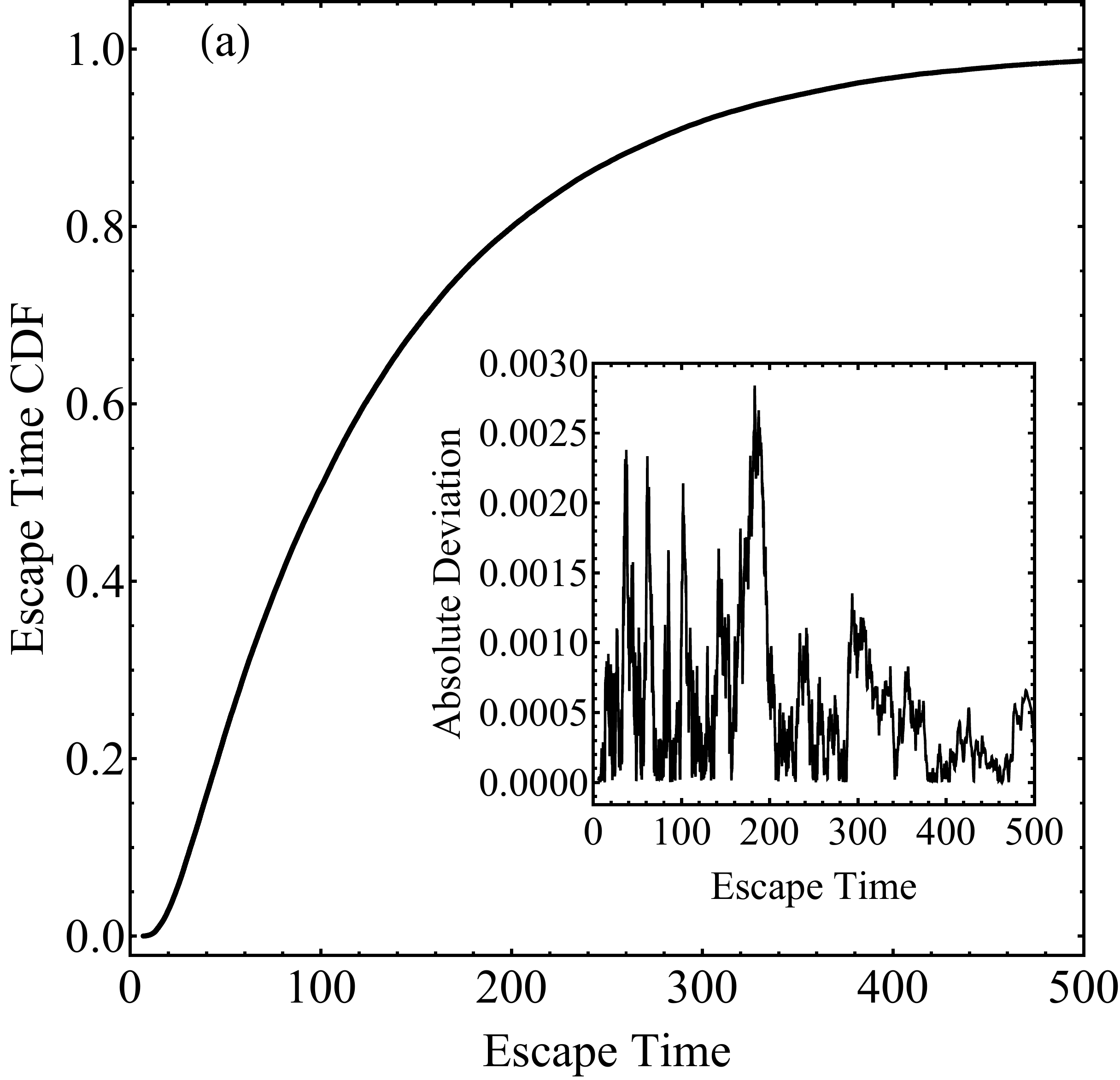}
\includegraphics [scale=0.099]{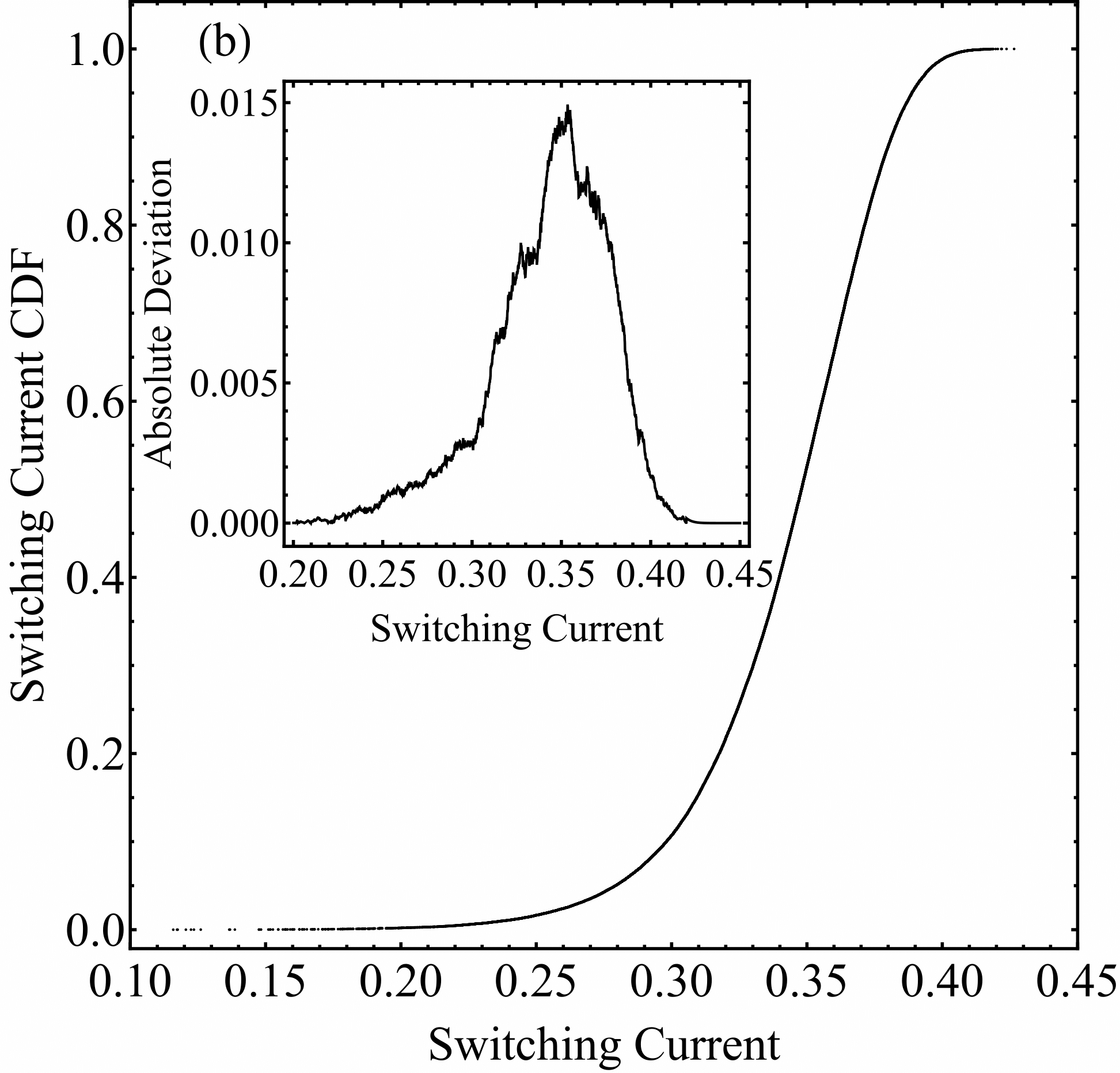}
\caption{Cumulative distributions of the escape time and switching current, and deviations from other reference models.
(a) Escape time cumulative distribution through Monte Carlo solution of  Eq.(\ref{JJn}). 
The numerical simulations are performed with CUDA Fortran following Algorithm \ref{algo:escape} 
and with ordinary GNU compiler in standard double precision arithmetic. 
On the scale of the figure the two  numerical solutions are the same, therefore the difference between the two results is displayed in the inset.
(b) Numerical simulations and theoretical predictions for the switching current cumulative distribution. 
The  numerical simulations have been performed with CUDA Fortran following Algorithm \ref{algo:switching}, while the theoretical predictions are based on Kramers adiabatic rate \cite{Fulton74}. 
On the scale of the figure the two  results are the same, therefore the difference is displayed in the inset.
Parameters of the simulations, reported in physical units,  are the same of a realistic device tested in Ref. \cite{Li07}: 
$R=250 \Omega$,  $C=88 \times 10^{-15}F$, $I_c = 0.748 \times 10^{-6}A$, $T=1.2K$, and the sweep rate is $200Hz$.
The SC empirical CDF  is computed using $N_{r} = 30\, 000$ realizations. 
The initial bias in each realizations is $\gamma = 0$.
The integration step $h = 10^{-4}$ corresponds to about $2 \times 10^{-4}$ of the natural oscillating period, that is given by the dimensionless plasma frequency $V_0^{1/2}$.}
\label{fig10}
\end{figure}

As a benchmark for the simulations we also employ the switching current distribution of the Josephson junctions.
We have preliminarily checked that CUDA simulations are statistically reliable (see Sect. \ref{Sec:Arrhenius}). 
The accuracy is further confirmed by the comparison of  simulations with ordinary double precision CPU  arithmetic and the GPU arithmetic.
{\hg 
For the sake of comparison, we have employed PGI Suite Compilers version 17.4, community edition, with Cuda Toolkit $8.0$ for both the CPU-only and the GPU version of the code. 
Routines have therefore been encoded without external libraries, such as those provided  by the Intel Math Kernel Library, that includes optimized implementation of Mersenne Twister random number generator.
}
The comparison is shown in  Fig. \ref{fig10}(a) as the escape times cumulative distributions computed for the same parameters for both GPU and CPU code.

The difference between the two distributions (shown in the inset) is  small and lies within the statistical fluctuations for this type of stochastic simulations.

Having tested the code, we now turn our attention to some physically interesting results.
The high speed simulations allow to investigate the adiabatic approximation used in \cite{Fulton74} that, neglecting non-equilibrium corrections to the Kramers' theory, gives a formula for the SC cumulative distribution only valid in the limit of vanishingly small bias ramp speed {\hg (for possible generalizations of the adiabatic approach, however overdamped,  see \cite{Pankratov00} } .
This formula can be compared  with the numerical result of  Fig. \ref{fig10}(b), that displays the computed cumulative distribution function of the JJ switching currents with the same parameters of the experiments reported in Ref. \cite{Li07}.
In particular,  to check  the theoretical approximations  we have performed simulations with a bias current step that  in normalized units correspond to about $10^{-12}$. 
It is shown that  simulated and theoretical SC cumulative distributions agree within $1.5 \%$, as displayed in the inset of Fig. \ref{fig10}(b).
This computation has required about  $1.5\times 10^5 sec \simeq 2 d$ on the GPU device.
Assuming an acceleration of $\simeq 400$ we estimate an execution time of about $5\times 10^8 sec \simeq 700 d$ on an ordinary CPU (e.g., Intel Core $i7-6700$). 
The direct numerical simulations of the experiments in the timescale of the laboratory electronics is therefore very demanding, for the different time scales between 
the JJ internal dynamics and the driving external electronics (see Sect. \ref{Model}).

%%%%%%%%%%%%%%%%%%%%%%%%%%%%%%%%%%%%%%%%%%%
%%%%%%%%%%%%%%%%%%%%%%%%%%%%%%%%%%%
\section{\label{Conclusions} Conclusions}

We have challenged the problem of an effective and fast algorithm for the mean first passage times with CUDA.
This is specially useful in the GPU simulations of escape times of Josephson junctions, where the intrinsic timescale of the system is extremely fast (up to the THz magnitude), while the electronic to control the escapes  from the zero voltage state is much slower (typically around or below 1KHz).

We have exploited the CUDA environment to compute in parallel several different realizations of the escape process  as schematically shown in Figs. \ref{fig:evolution} 
and \ref{fig:streaming}.
Each parallel thread uses a pseudorandom number generator with a different seed.

%%%%%%%%%%%%___FIG___ comparison_____%%%%%%%
%%%%%%%%%%END FIGURE%%%%%%%%%%%%%%%%%%%%%%%%%%

The efficiency of the proposed algorithm reaches accelerations of about $400$ respect to standard Intel Core $i7-6700$ host processors. 
The scaling of the performances with the rising of the number of threads has ben elucidated in Figs. \ref{fig:exec} and \ref{fig:exec_time_comparison}. 
From the data, it is clear that: 
i) the choice of random generator is crucial, it should be quick and low memory demanding; 
ii) the load for each thread should be high enough to avoid that fluctuations dominate and reduce the efficiency of parallelism{\hg;
iii) in the running regime (that, however, is not essential for switching current distributions) the argument of the sin, namely $\varphi$, grows to a level where the single precision could be not enough }.
With this care the abovementioned efficiency  applied to the benchmark Josephson junctions proved fast enough to challenge realistic experiments \cite{Li07}.
The numerical experiments have reproduced practical set-ups without extrapolation formulas and could be useful, for instance, to investigate the thermal-quantum transition process.
In the same field we suggest that CUDA environment might be also relevant for arrays of Josephson junctions, possibly coupled to a resonator \cite{Yamapi14} or for voltage standards \cite{Behr12}.

%%%%%%%%%%%%%%%%%%%%%%%%%%%%%%%%%%%%%%%%%%%%%%%%%%%%%%%%%%%%%%%%%%%%%%%%%%%%%%%%%%%%

\section*{Acknowledgements}

We thank I. M. Pinto and S. Pagano for stimulating discussions.
%
% We acknowledge financial support from 
%
VP acknowledges INFN, Sezione di Napoli (Italy) for partial financial support .

\appendix

\section*{Appendix A. Parallelism Statistical Model}
\setcounter{equation}{0}\renewcommand{\theequation}{A.\arabic{equation}}

We propose a simple probabilistic model for the parallel computation of escape times with GPU in CUDA environment that is at the basis of the performances of Fig. \ref{fig:speedup}.

The execution time of Algorithm \ref{algo:escape}  for each parallel process is directly proportional to the random escape time.
The overall average execution time random variable $\langle E_t \rangle$  is determined by the sum of the individual escapes $t_i$:
\beq
E_t  = \sum_{i=1}^{N_r}t_{i}.
\label{exectime}
\eeq
The variable $t_i$ can be approximated with an exponential random deviate (although this is not exact, for the presence of an inertial short time prior to escapes over the barrier) identically distributed with an appropriated time scaling. 
The random variable (\ref{exectime}) is distributed as the sum of iid exponentials' and the resulting distribution is an $Erlang(\cdot)$ function \cite{Evans00}
\beq
E_t \sim Erlang(N_r,1).
\eeq
Defining  $N_L=N_r/N_{tpb}$ the number of  runs  assigned to each parallel thread the execution time random variable of the parallel algorithm is 
\beq
E_p = \max(E_t^{(1)},E_t^{(2)},...,E_t^{(N_{tpb})}).
\label{maxtime}
\eeq
%and  is a random variable distributed as the maximum of the  execution time of the independent processes.
We define the parallel execution time as the average $\langle  E_p \rangle$.
From (\ref{maxtime}) it follows that the CDF of the random variable  $E_p$ reads
\beq
CDF[E_p,x]= CDF[Erlang(N_L,1),x]^{N_{tpb}}
\label{eq:parexec}
\eeq
by Eq. (\ref{eq:parexec})  is possible to compute in closed form all the moments,  in particular the parallel execution time $\langle  E_p \rangle$.
If one defines the average speedup as the ratio 
\beq
S = \frac{ \langle E_t \rangle}{\langle  E_p \rangle },
\label{eq:speedup}
\eeq
a perfect parallel algorithm entails: $S  = N_{tpb}$. 
To display a quantity that is independent of the time scaling used in (\ref{exectime}) it is convenient to define the speedup efficiency as the ratio $S /N_{tpb}$.
The theoretical estimate (\ref{eq:speedup})  for the speedup is the basis for the results displayed in Fig. \ref{fig:speedup} and discussed in Sect. \ref{parallelism}.


\section*{References}
\begin{thebibliography}{00}



\bibitem{Risken89} H. Risken, {\it The Fokker-Planck Equation.} (Springer-Verlag, 1984)
%\bibitem{Mannella} Paper on problems with thresholds in stochastic systems
\bibitem{Pollak16} E. ~Pollak and R. ~Ianconescu, J. Phys. Chem. A,  {\bf 120}, 3155 (2016).
\bibitem{Mazo10} J. J. ~Mazo, F. ~Naranjo, and D. ~Zueco, Phys. Rev. B {\bf 82},  094505 (2010).


\bibitem{Barone82} A.~Barone, G.~Patern\`o, {\em ``Physics and Applications of the Josephson Effect''}, (John Wiley \& Sons, 1982).
\bibitem{DevoretBook} M. H.~Devoret, {\em ``Quantum Fluctuations in Electrical Circuits''}, S. Reynaud, E. Giacobino and J. Zinn-Justin, eds., Les Houches, Session LXIII, 1995, Elsevier (1997).
\bibitem{Pierro16}  V. Pierro and G. Filatrella, Phys. Rev. A {\bf 94}, 042116 (2016).
%Nonideal quantum measurement effects on the switching-current distribution of Josephson junctions
\bibitem{Pekola04} J. P. Pekola , Phys. Rev. Lett. {\bf 93}, {206601} (2004).
% Josephson Junction as a Detector of Poissonian Charge Injection},
\bibitem{Filatrella2010} G. ~Filatrella and V. ~Pierro, Phys. Rev. E {\bf 82}, 046712 (2010).
\bibitem{Addesso2012} P. ~Addesso, G. ~Filatrella, and V. ~Pierro, Phys. Rev. E {\bf 85}, 016708 (2012).
\bibitem{Gross13} B. Gross, J. Yuan, D. Y. An, M. Y. Li, N. Kinev, X. J. Zhou, M. Ji, Y. Huang, T. Hatano, R. G. Mints, V. P. Koshelets, P. H. Wu, H. B. Wang, D. Koelle, and R. Kleiner
%"Modeling the linewidth dependence of coherent terahertz emission from intrinsic Josephson junction stacks in the hot-spot regime"
Phys. Rev. B {\bf 88}, 014524 (2013).




\bibitem{Shukrinov17}Yu. M. Shukrinov, I. R. Rahmonov, K. V. Kulikov, A. E. Botha,
A. Plecenik, P. Seidel and W. Nawrocki,
%"Modeling of LC-shunted intrinsic Josephson junctions in high-Tc superconductors"
Supercond. Sci. Technol. {\bf 30},  024006 (2017).

\bibitem{Galin15} M. A. Galin, A. M. Klushin, V. V. Kurin, S. V. Seliverstov, M. I. Finkel, G- N- Goltsman, F. M\"uller, T. Scheller and A. D. Semenov,
%"Towards local oscillators based on arrays of niobium Josephson junctions",
Supercond. Sci. Technol. {\bf 28}, 055002 (2015).
\bibitem{Freund09} L. B.  ~Freund, Proc. Natl. Acad. Sci. U.S.A. {\bf 106}, 8818 (2009). 
\bibitem{Mazo13} J. J. ~Mazo, F. ~Naranjo, and D. ~Zueco,  J. of Chem. Phys. {\bf 138}, 104105 (2013).
%doi / 10.1073 / pnas.0903003106
%Characterizing the resistance generated by a molecular bond as it is forcibly separated


\bibitem{Smirnov10}
%  title = {Influence of the size of uniaxial magnetic nanoparticle on the reliability of high-speed switching},
 A. A. Smirnov, and A. L. Pankratov, Phys. Rev. B {\bf 82}, 132405 (2010).
%  month = {Oct},
%  publisher = {American Physical Society},
%  doi = {10.1103/PhysRevB.82.132405},
%  url = {https://link.aps.org/doi/10.1103/PhysRevB.82.132405}


\bibitem{Januszewski10} M. Januszewski, M. Kostur,  Comput. Phys. Commun. {\bf 181}, 183 (2010).

\bibitem{Fedorov09} K. G. Fedorov and A. L. Pankratov, Phys. Rev. Lett. {\bf 103}, 260601 (2009) .

\bibitem{graham85} R. Graham and T. T\'el, Phys. Rev. A {\bf 31}, 1109 (1985).
\bibitem{kautz94} R.L. Kautz, J. Appl. Phys. {\bf 76}, 5538 (1994). 

\bibitem{Artemiev11}
%“Numerical solution of stochastic differential equations on Supercomputers”, 
S.S. Artemiev and V.D. Korneev, Numer. Analys. Appl. {\bf 4}, 1 (2011).
% https://doi.org/10.1134/S1995423911010010.


\bibitem{Buttiker83} M. B\"uttiker, E.P. Harris, and R. Landauer, Phys. Rev. B {\bf 28}, 1268 (1983).
%  title = {Thermal activation in extremely underdamped Josephson-junction circuits},
\bibitem{Augello09} G. Augello, D. Valenti, A.L. Pankratov, and B. Spagnolo, Eur. Phys. B {\bf 70}, 145  (2009).
\bibitem{Groenbeck04} N. Gr\o nbech-Jensen, M. G. Castellano, F. Chiarello, M. Cirillo, C. Cosmelli, L. V. Filippenko, R. Russo, and G. Torrioli,  Phys. Rev. Lett. {\bf 93}, 107002 (2004).
%"Microwave-Induced Thermal Escape in Josephson Junctions"
\bibitem{Groenbeck05} N. Gr\o nbech-Jensen and M. Cirillo, Phys. Rev. Lett.  {\bf 95}, 067001 (2005).
%"Rabi-Type Oscillations in a Classical Josephson Junction,"
%1
\bibitem{Martinis87}J. M. Martinis, M. H. Devoret, and J. Clarke, Phys. Rev. B {\bf 35}, 4682 (1987).
%2
\bibitem{Shnirman97} A. Shnirman, E. Ben-Jacob, and B. Malomed, Phys. Rev. B,  {\bf 56}, {14677} (1997).
%% esempio di tunnel mesoscopico   title = {Tunneling and resonant tunneling of fluxons in a long Josephson junction},
\bibitem{Martinis02} J. M. Martinis, S. Nam,  J. Aumentado, and C. Urbina, Phys. Rev. Lett. {\bf 89}, 117901 (2002).
%3
% ``Rabi Oscillations in a Large Josephson-Junction Qubit ``, 
\bibitem{Wallraff03b} A. Wallraff, A. Lukashenko, J. Lisenfeld, A. Kemp, M. V. Fistul, Y. Koval, and A. V. Ustinov, Nature {\bf 425}, 155 (2003).
% Quantum dynamics of a single vortex
%4
\bibitem{Price10} A. N. Price, A. Kemp, D.R. Gulevich, F.V. Kusmartsev, and A.V. Ustinov, {Phys. Rev. B}, {\bf 81}, {014506} (2010).
% {Vortex qubit based on an annular Josephson junction containing a microshort},
\bibitem{Coskun12}U. C. Coskun, M. Brenner, T. Hymel, V. Vakaryuk, A. Levchenko, and A. Bezryadin,  {Phys.  Rev. Lett.} {\bf  108},  097003 (2012).
%"Distribution of Supercurrent Switching in Graphene under the Proximity Effect," 
%6
\bibitem{Massarotti15} D. Massarotti, A. Pal, G. Rotoli, L. Longobardi, M.G. Blamire, and F. Tafuri, Nature Comm. {\bf 6}, 7376 (2015).
\bibitem{Makhlin01}Y. Makhlin, G. Sch\"on, and A. Shnirman,  Rev. Mod. Phys. {\bf 73}, 357 (2001).
%Quantum-state engineering with Josephson-junction devices, 
\bibitem{Blackburn16}J. A. Blackburn, M. Cirillo, N. Grønbech-Jensen, Physics Reports {\bf 611}, 1 (2016).
%A survey of classical and quantum interpretations of experiments on Josephson junctions at very low temperatures
\bibitem{EulerInt} P. E. Kloeden, E. Platen and H. Schurz, {\it Numerical Solution of SDE Through Computer Experiments} (Springer-Verlag, 1994). 

\bibitem{Pankratov97}A.L. Pankratov   Phys. Lett. A {\bf 234},  329 (1997).
\bibitem{Malakhov96} A.N. Malakhov , A.L. Pankratov,  Physica C {\bf 269},  46 (1996).


\bibitem{Numrecip} W. H. Press, S. A. Teukolsky, W. T. Vetterling, and B. P. Flannery, {\it Numerical Recipes} (Cambridge University Press, 1995), Vol. I, Ch. 19.
\bibitem{Matsumoto98} M. Matsumoto, and T. Nishimura,  ACM Trans. Model. Comput. Simul. {\bf 8}, 3 (1998).
%la pagina controlla Mersenne twister: a 623-dimensionally equidistributed uniform pseudo-random number generator. 
\bibitem{Fulton74} T. A. Fulton and L. N. Dunkleberger, Phys Rev B {\bf 9}, 4760 (1974).
\urlstyle{rm}
\bibitem{floating} \url{http://docs.nvidia.com/cuda/floating-point/index.html}, consulted on January 8th, 2018.


\bibitem{Pankratov00} A. L. Pankratov and M. Salerno, Phys Rev E {\bf 61}, 1206 (2000); Phys. Lett. A {\bf 273}, 162 (2000).


\bibitem{Li07} Shao-Xiong Li, Wei Qiu, Siyuan Han, Y. F. Wei, X. B. Zhu, C. Z. Gu, S. P. Zhao, and H. B. Wang, Phys. Rev. Lett. {\bf 99}, 037002 (2007).
%Observation of Macroscopic Quantum Tunneling in a Single Bi$_2$Sr$_2$CaCu$_2$O$_{8-\delta}$ Surface Intrinsic Josephson Junction


\bibitem{Yamapi14}R. Yamapi and G. Filatrella,  Phys. Rev. E {\bf 89}, 052905 (2014).
%Noise effects on a birhythmic Josephson junction coupled to a resonator
\bibitem{Behr12}R. Behr, O. Kieler, J. Kohlmann, F. M\"uller, and L. Palafox,  Meas. Sci. Technol. {\bf 23}, 124002 (2012).
%  (19pp), doi:10.1088/0957-0233/23/12/124002  ,
%Development and metrological applications of Josephson arrays at PTB
\bibitem{Evans00} M. Evans, N. Hastings, and B. Peacock, {\it Statistical Distributions},  4th Ed. (John Wiley \& Sons, 2011)  Ch. 15, p. 84.


\end{thebibliography}


\end{document}